\documentclass[twocolumn,superscriptaddress,prl]{revtex4-2}
\usepackage{amsmath}
\usepackage{graphicx}
\usepackage{xcolor} 
\usepackage[normalem]{ulem} 

\newcommand{\stkout}[1]{\ifmmode\text{\sout{\ensuremath{#1}}}\else\sout{#1}\fi}
\newcommand{\md}{\mathrm d}

\newcommand{\mcQ}{\mathcal Q} 
\newcommand{\mcW}{\mathcal W} 
\newcommand{\mcU}{\mathcal U} 
\newcommand{\mcV}{\mathcal V}

\DeclareMathOperator{\sech}{sech}

\begin{document}

\title{Connected Network Model for the Mechanical Loss of Amorphous Materials}

\author{Steven Blaber}
\email{steven.blaber@ubc.ca}
\affiliation{Dept.~of Physics and Astronomy and Stewart Blusson Quantum Matter Institute, University of British Columbia, Vancouver, British Columbia V6T 1Z1, Canada}
\author{Daniel Bruns}
\affiliation{Dept.~of Physics and Astronomy and Stewart Blusson Quantum Matter Institute, University of British Columbia, Vancouver, British Columbia V6T 1Z1, Canada}
\author{Jörg Rottler}
\email{jrottler@physics.ubc.ca}
\affiliation{Dept.~of Physics and Astronomy and Stewart Blusson Quantum Matter Institute, University of British Columbia, Vancouver, British Columbia V6T 1Z1, Canada}

\begin{abstract}
Dissipation in amorphous solids at low frequencies is commonly attributed to activated transitions of isolated two-level systems (TLS) that come in resonance with elastic or electric fields. Materials with low mechanical or dielectric loss are urgently needed for applications in gravitational wave detection, high precision sensors, and quantum computing. Using atomistic modeling of amorphous silicon and titanium dioxide, we find that their energy landscape is better represented by a connected network of inherent structures than a collection of isolated TLS. Each \emph{connection} is a single energy barrier between two minima, and a network is \emph{connected} if all states (minima) can be reached from any given state. Motivated by this observation, we develop an analytically tractable theory for mechanical loss of the full network from a nonequilibrium thermodynamic perspective. We demonstrate that the connectivity of the network introduces new mechanisms that can both reduce low frequency dissipation through additional low energy relaxation pathways, and increase dissipation through a broad distribution of energy minima. As a result, the connected network model predicts mechanical loss with distinct frequency profiles compared to the isolated TLS model. This not only calls into question the validity of the TLS model, but also gives us many new avenues and properties to analyze for the targeted design of low mechanical loss materials.
\end{abstract}
\date{\today}

\maketitle

\section{Introduction}

Amorphous materials are key components in many fast developing technologies, yet the performance of these devices is often severely limited by internal friction. For instance, dielectric loss in the aluminum oxide layers of solid state superconducting qubits limits their lifetime and thus qubit coherence \cite{martinis2005decoherence}. In Laser Interferometer Gravitational Wave Detectors (GWD), it is instead mechanical dissipation in the amorphous oxide mirror coatings that is currently a significant limiting factor for the sensitivity of the instrument \cite{steinlechner2018,vajente2021low}. Finding strategies to reduce loss mechanisms is thus paramount. Mechanical loss also plays an important role in the field of cavity optomechanics~\cite{aspelmeyer2014}, which has applications to quantum memory, high precision sensors, and quantum transducers.

The physical origin of energy dissipation in amorphous films is ascribed to the presence of two-level systems (TLS), whose thermally activated transitions come in resonance with a mechanical or electromagnetic wave \cite{damart2018}. The TLS represent atomic level transitions within amorphous solids; for example, amorphous silicon (a-Si) may undergo bond hopping transitions~\cite{levesque2022}. The two distinct atomic configurations may be associated with pairs of stable minima (inherent structures) in an energy landscape connected by a transition path. External fields drive the TLS out of equilibrium, resulting in internal friction upon relaxation~\cite{phillips1987,damart2018,berthier2023}. Mechanical and dielectric loss are caused by small displacements of atoms constituting the TLS, with mechanical waves coupling to the TLS in mechanical loss and electromagnetic waves coupling to the TLS' dipole moment in dielectric loss.

Although TLS can interact with each other through long range fields, they are usually taken to be spatially independent entities. As a result, the total dissipation is just the sum of the contributions from each TLS~\cite{phillips1987,damart2018,berthier2023}. This assumption of independence, however, is in sharp contention with the well established notion of amorphous solids being described by a rugged, complex, and high-dimensional energy landscape~\cite{berthier2023,raza2015}. The TLS form part of this energy landscape, where in general there are more than just one transition path out of a particular local energy minimum.

We explore the consequences of accounting for the rugged and high-dimensional energy landscape through a \emph{connected network} model. A connected network is
a discrete-state representation of the high dimensional energy landscape, with each node representing an energy minimum and each connection (edge) a single energy barrier. In a connected network, every state is accessible from any given state, giving a consistent thermodynamic description of the energy landscape. Our focus lies on classical thermal transitions as they relate to the mechanical loss near room temperature, and we take a non-equilibrium thermodynamic perspective~\cite{Seifert2012}: modeling the heat dissipated from a driven process using master equation dynamics~\cite{Gardiner} of the discrete state network. In the limit of linear response, we arrive at a rigorous generalization of the 50 year old single TLS model that accounts for the additional relaxation paths afforded by the connectivity of the network. Through simple model systems and full atomistic modeling, we reveal mechanisms by which activated transitions in a larger connected network of energy minima can significantly reduce mechanical loss in the frequency band of $10^1-10^4\,$Hz where ground-based gravitational wave detection takes place. 

\begin{figure*}
        \includegraphics[width=\textwidth]{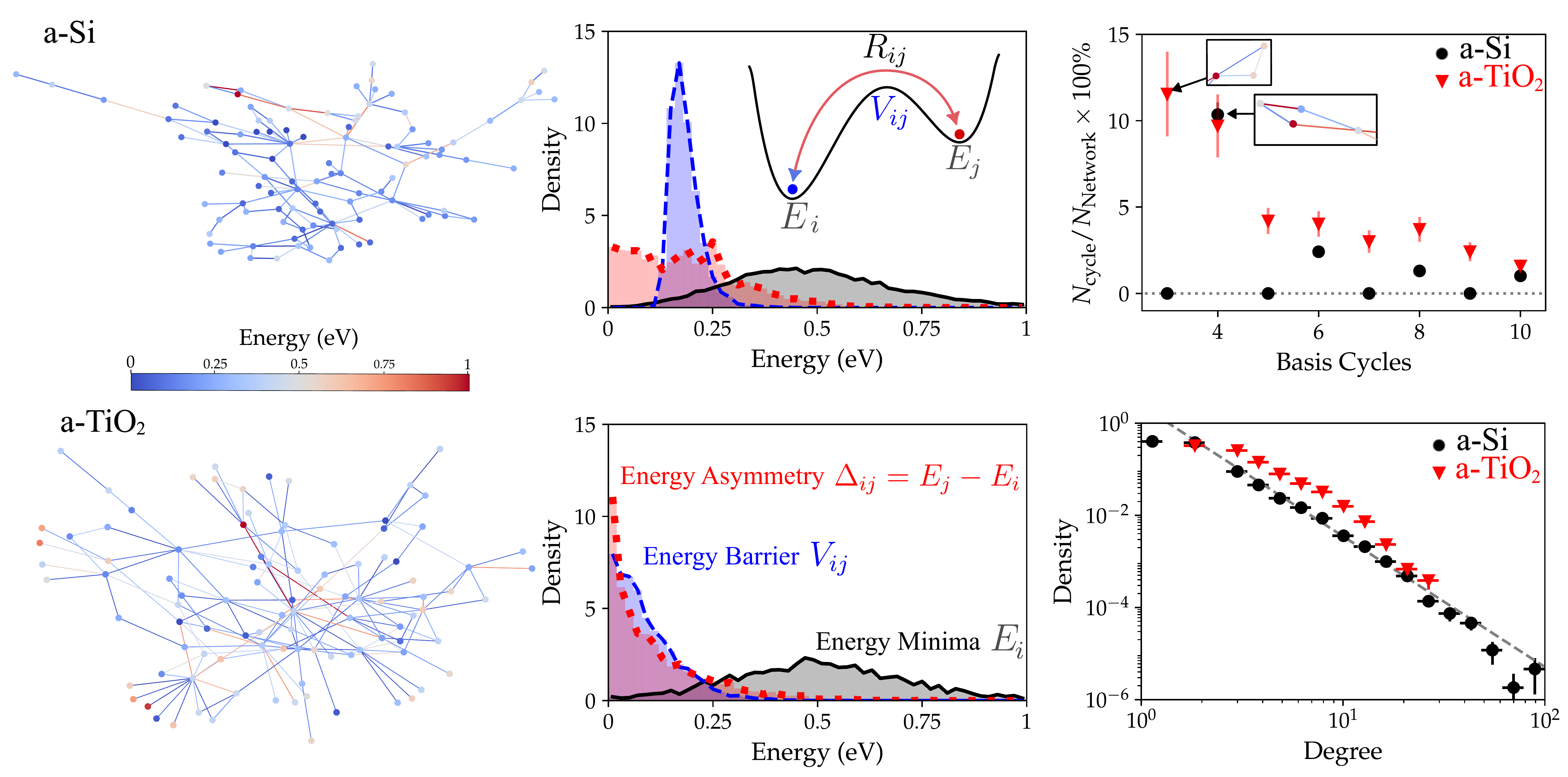}
	\caption{Left: First $100$ nodes of the connected network of inherent structures observed in samples of amorphous silicon (a-Si) and amorphous titanium dioxide (a-TiO$_{2}$). The color of the dots and lines correspond to the energy of the minima $E_{i}$ and the average barrier height $V_{ij} - (E_i+E_j)/2$ respectively. Middle: Distributions of energy minima $E_{i}$ (black solid curve), magnitude of energy asymmetries $\Delta_{ij} = E_j-E_i$ (red dotted), and barrier heights $V_{ij}$ (blue dashed) between inherent structures ($i$,$j$) for a-Si (top) and a-TiO$_{2}$ (bottom). The energetics for a pair of connected inherent structures is shown schematically in the inset. Right:  Fraction of the number of basis cycles and probability distribution of the degree (number of connections) of nodes in the network with the dashed line a linear least squares fit to the a-Si data with a slope (scaling exponent) of $-2.84\pm 0.04$. Error bars in the degree distribution are the standard error of the mean from logarithmically binned averages. Basis cycles are the minimal set of closed loops in the network, with examples of a three-state cycle in a-TiO$_{2}$ and a four-state cycle in a-Si shown in the inset.}
	\label{fig: TLS_Network}
\end{figure*}

\section{Energy landscapes from atomistic simulations}

We begin by exploring the energy landscapes of \emph{in-silico} samples of a-Si and a-TiO$_{2}$ as described in section~\ref{Sect: Materials and Methods}. Through thermal molecular dynamics search trajectories, we find inherent structures (stable minima of the energy landscape) connected by thermodynamic pathways determined through nudged elastic band (NEB) calculations. Representing energy minima as nodes and thermodynamic transitions as edges, the energy landscapes form sparsely connected networks (Fig.~\ref{fig: TLS_Network}, left panel). Each connected pair of nodes would be considered an individual two-level system within the TLS model (Fig.~\ref{fig: TLS_Network}, middle panel inset). Although only the first 100 nodes are shown for clarity, the observed networks often contain thousands of inherent structures (nodes).  In a-Si, the percentage of the network that is connected is estimated to be $\sim 0.05\%$, while in a-TiO$_{2}$ the connectivity of the largest networks is $\sim0.14\%$. Although the energy landscape search via a random walk is not guaranteed to explore all edges or inherent structures, we find these connectivity values robust and converged within our numerical protocol (see Supplemental Material~\ref{app: convtests}).

The distributions of both the energy minima and the energy barriers that link them are broad (middle panel), reflecting the structural disorder. Locally, we find that nearby energy minima often form closed loops or cycles (right panel). The degree distribution of these networks follows a power law, indicating that the networks are scale free. Similar scale-free networks are found in Lennard-Jones clusters~\cite{doye2002}, while the energy landscape picture has been invoked to explain numerous phenomena associated with the glass transition ~\cite{raza2015,angelani1998connected,banerjee2012characterization}.

Traditionally, each transition between inherent structures is treated as a new independent TLS~\cite{damart2017}. Since the inherent structure describes a minimal energy configuration of the entire molecular system, in principle the configurations of $N$ TLS can result in $2^N$ inherent structures. The network description of the TLS model with $N$ pairs of states would form a $2^N$ state hypercubic network with delta function degree distribution $\delta({\rm degree}-N)$ and only 4-state basis cycles. The mechanical loss of such a network is identical to the TLS model only if the transitions and their response to mechanical perturbations are independent from each other. For example, two spatially separated bond-hop transitions that can oscillate independently would form a $2^2=4$ state network with one four-state cycle (Supplemental Material~\ref{app: TLS derivation}).

Since the energy landscape of amorphous materials form connected networks instead of independent pairs of transitions as assumed by the TLS model, we now explore both theoretically and numerically the implications of connectivity on mechanical loss. Loops and cycles form topologically distinct structures that cannot be accounted for by single TLSs, and the broad energy distributions from the structural disorder will require proper treatment in terms of the global energy scale.

\section{Thermal Connected Network Model}

\subsection{Thermodynamics}
Mechanical loss of a material relates to the decay of acoustic vibrations due to internal energy dissipation. Experimentally, mechanical loss can be estimated by driving a coated cantilever at a resonant frequency of the cantilever~\cite{murray2015}. When the mechanical driving is turned off, the subsequent decay of the mechanical wave (attenuation or dampening) is measured and can be used to determine the mechanical loss. From a theoretical viewpoint, the attenuation of mechanical waves results from energy dissipated as heat $\mcQ_{\rm cycle}$ in each oscillation, which is compared to the stored elastic energy through the inverse \emph{quality factor}
\begin{align}
	Q^{-1} = \frac{1}{2\pi}\frac{\mcQ_{\rm cycle}}{{\rm energy~ stored}} \ .
	\label{eq: quality factor definition}
\end{align}
This relation allows us to relate the quality factor to the dissipated heat, which can then be used to estimate the quality factor based on material properties. The inverse quality factor is what we refer to as \emph{mechanical loss}.

We separate the total energy of the system into the elastic energy of the acoustic wave of frequency $\omega$ traveling through the material and the internal energy of the connected network of inherent structure as $\mcU_{\rm tot} = \mcU_{\rm elastic} + \mcU_{\rm CN}$. Assuming an isotropic and linear elastic material with oscillating strain $\epsilon(t) = \epsilon_0\sin(\omega t)$, the elastic work is
\begin{align}
	\mcW_{\rm elastic}(t) = \mcV C \epsilon_0^2\sin^2(\omega t) \ ,
	\label{eq: elastic energy} 
\end{align}
for volume $\mcV$ and elastic modulus $C$ (longitudinal or shear).   The macroscopic energy input of the entire material oscillating under strain averaged over one cycle is $\mcV C \epsilon_0^2/2$.

The strain couples to the energy of the inherent structures $\boldsymbol{E}$ as 
\begin{align}
	\boldsymbol{E}(t) = \boldsymbol{E}(0) + \left(\alpha\boldsymbol{1} + \frac{\epsilon_0\gamma_0}{2}\boldsymbol{\Gamma}\right) \sin(\omega t) \ ,
	\label{eq: Time Dependent Energy}
\end{align}
where $\boldsymbol{1}$ is a vector of ones and throughout a bold symbol represent a vector with elements spanning the inherent structures; i.e. $E_{i}$ is the energy of inherent structure $i$. The coupling between the inherent structure and applied strain is separated into a constant term $\alpha$ and an inherent structure dependent term $\epsilon_0\gamma_0\boldsymbol{\Gamma}/2$, with strain amplitude $\epsilon_0\gamma_0$ and dimensionless inherent structure coupling $\boldsymbol{\Gamma}$. The constant term will not affect the dynamics since it will affect all energies equally. The average energy of the network is the number of inherent structures $N$ times the average energy per inherent structure
\begin{align}
	\mcU_{\rm CN}(t) = N\boldsymbol{P}_{t}\cdot \boldsymbol{E}(t) \ ,
\end{align}
with $\boldsymbol{P}_{t}$ the time dependent inherent structure occupation probability.

The rate of change in energy gives the first law of thermodynamics 
\begin{align}
	\dot{\mcU}(t) = \dot{\mcW}(t) + \dot{\mcQ}(t) \ ,
\end{align}
with the work resulting from changes in energy
\begin{align}
	\dot{\mcW} = \dot{\mcW}_{\rm elastic} + N\boldsymbol{P}_t\cdot \dot{\boldsymbol{E}}(t) \ ,
	\label{eq: Work}
\end{align}
and the heat from the time dependent probabilities
\begin{align}
	\dot{\mcQ} = N\dot{\boldsymbol{P}}_t \cdot \boldsymbol{E}(t) \ .
	\label{eq: Heat}
\end{align}
Throughout, a dot denotes the rate of change with respect to time. The work and heat produced in a cyclic process are $\mcW_{\rm cycle} = \int_{\rm cycle} \md t~ \dot{\mcW}$ and $\mcQ_{\rm cycle} = \int_{\rm cycle} \md t~\dot{\mcQ}$. For a cyclic process in a periodic steady-state $\Delta \mcU_{\rm cycle} = 0$ so $\mcQ_{\rm cycle} = -\mcW_{\rm cycle}$.

\subsection{Dynamics}
We describe the dynamics of the system in the energy landscape by a \emph{master equation}, which is a discrete state model that describes conservation of probabilities as they transition between states~\cite{Gardiner}
\begin{align}
	\frac{\md \boldsymbol{P}_{t}}{\md t} = R(t)\boldsymbol{P}_{t} \ ,
	\label{eq: ME General}
\end{align}
with a transition rate matrix $R(t)$. Such an approach has previously been employed to elucidate various facets of slow dynamics in supercooled liquids, such as stretched exponential relaxation, the breakdown of the Stokes-Einstein relation \cite{angelani1998connected}, and the crossover from fragile to strong glass forming behavior \cite{banerjee2012characterization}.  

For a system at inverse temperature $\beta \equiv (k_{\rm B} T)^{-1}$ with temperature $T$ and Boltzmann constant $k_{\rm B}$, the transition rates are expressed as Arrhenius rates in terms of the energy barriers $V_{ij}$ and energy levels $E_i$ of the inherent structures $i$ and $j$. The transition rate matrix has elements
\begin{align}
	&R_{ij} =k_{ij}e^{\beta E_{j}}\left[e^{-\beta V_{ij}}(1-\delta_{ij}) - \delta_{ij}\sum_{\ell\neq j}e^{-\beta V_{j\ell}}\right] \ .
	\label{eq: transition matrix}
\end{align}
Due to the exponential dependence, the transition rates are dominated by the barrier heights $V_{ij}$ rather than the bare transition rates $k_{ij}$~\cite{trinastic2016}. For simplicity, we assume equal bare transition rates $k_{ij} = k_{0}$ for all transitions in our numerical calculations, an assumption made in a previous study~\cite{levesque2022} and 
supported by atomistic simulations of amorphous silicon~\cite{valiquette2003}.

If the transition rate matrix $R$ is time independent, then~\eqref{eq: ME General} has the solution
\begin{align}
    \boldsymbol{P}_{t} &= e^{R t} \boldsymbol{P}_{0} \ , \nonumber \\
    &= Me^{\Lambda t} M^{-1}\boldsymbol{P}_{0} \ ,
\end{align}
for initial condition $\boldsymbol{P}_{t} = \boldsymbol{P}_{0}$, and matrix exponential $e^{Rt}$. In the second line we have assumed that $R$ can be diagonalized as $R = M\Lambda M^{-1}$, with $M$ the matrix with columns the eigenvectors of $R$, and $\Lambda$ the corresponding eigenvalue matrix with diagonal elements the eigenvalues of $R$. For a connected network there is exactly one zero eigenvalue corresponding to the equilibrium distribution
\begin{align}
    \boldsymbol{P}^{\rm eq} = \frac{e^{-\beta \boldsymbol{E}}}{Z} \ ,
    \label{eq: equilibrium probability}
\end{align}
with the partition function $Z\equiv \sum_{i}\exp(-\beta E_{i})$. The remaining eigenvalues are negative and represent relaxation towards equilibrium.

If $R$ is time dependent, then the formal solution to~\eqref{eq: ME General} is
\begin{align}
    \boldsymbol{P}_{t} = \mathcal{T}\exp\left[\int_{0}^{t}\md t' R(t')\right]\boldsymbol{P}_{0} \ ,
\end{align}
where $\mathcal{T}\exp[\cdots]$ is the time ordered matrix exponential. Although useful for our understanding, the formal solution does not lend itself well to calculating the time-dependent probability $\boldsymbol{P}_{t}$, and so we turn to approximate methods.

\subsection{Linear Response (Weak Limit)}
Although the equations of the previous section (Eq.~\eqref{eq: Heat} and Eq.~\eqref{eq: ME General}) are sufficient to determine the energy dissipation and hence the mechanical loss of the system, considerable simplification can be made if we assume the amplitude of oscillations are small. This is consistent with the small amplitude assumption made in the TLS model~\cite{damart2018} and should be sufficient for applications to GWD where the magnitude of vibrations in the system are extremely small~\cite{steinlechner2018}.

For weak perturbations (of amplitude $\epsilon$) the probability distribution remains near its initial equilibrium state as
\begin{align}
    \boldsymbol{P}_{t} \approx \boldsymbol{P}^{\rm eq}_{0} + \boldsymbol{P}^{\rm weak}_{t} \ ,
    \label{eq: ME approx}
\end{align}
where the deviation from the initial equilibrium $\boldsymbol{P}^{\rm weak}_{t}$ is assumed to be small (of order $\epsilon$). In addition, the transition rate matrix also experiences a small deviation from its initial rates
\begin{align}
    R(t) \approx R_{0} + R^{\rm weak}(t) \ .
\end{align}
The second term $R^{\rm weak}(t)$ is assumed to be small (of order $\epsilon$).

Substituting both of the above into~\eqref{eq: ME General} we have
\begin{align}
    \frac{\md\boldsymbol{P}^{\rm weak}_{t}}{\md t} &= \left[R_{0} + R^{\rm weak}(t)\right]\left(\boldsymbol{P}^{\rm eq}_{0} + \boldsymbol{P}^{\rm weak}_{t}\right) \ , \nonumber \\
    &\approx R_{0} \boldsymbol{P}^{\rm weak}_{t} + R^{\rm weak}(t) \boldsymbol{P}^{\rm eq}_{0} \ .
\end{align}
In the first line we used that $ \boldsymbol{P}^{\rm eq}_{0}$ is time independent, and in the second line dropped terms of order $\epsilon^2$. In the steady state this has the solution
\begin{align}
     \boldsymbol{P}^{\rm weak}_{t} \approx \int_0^{t}\md t' e^{R_{0} t' }R^{\rm weak}(t') \boldsymbol{P}^{\rm eq}_{0} \ .
     \label{eq: weak}
\end{align}
For the specific case of oscillating energy minima according to~\eqref{eq: Time Dependent Energy} the solution is
\begin{align}
	\boldsymbol{P}^{{ \rm weak}}_t &=-\frac{\beta\epsilon_0\gamma_0}{2}\left[\boldsymbol{A} \sin (\omega t)+\boldsymbol{B} \cos (\omega t)\right] \label{eq: Probability approx} \\
	A_{i}&\equiv \sum_{jk} M_{ij}\frac{1}{1+(\omega\tau_{j})^2}M_{jk}^{-1} {\Gamma}_k P^{\rm eq}_k \nonumber\\
	B_i &\equiv \sum_{jk} M_{ij}\frac{\omega\tau_{j}}{1+(\omega\tau_{j})^2}M_{jk}^{-1} {\Gamma}_k P^{\rm eq}_k \ . \nonumber
\end{align}
The relaxation times $\tau_{j}$ are related to the eigenvalues $\lambda_{j}$ of $R_{0}$ as $\tau_{j} \equiv \lambda_{j}^{-1}$ with corresponding eigenvectors forming the columns of the eigenvector matrix $M$.

Substituting Eqs.~\eqref{eq: ME approx} and Eq.~\eqref{eq: Probability approx} into Eq.~\eqref{eq: Heat}, then integrating over one cycle, we have
\begin{align}
	\mcQ_{\rm CN} = \frac{\beta \pi N\epsilon_0^2\gamma_{0}^2}{4}\sum_{i,j,\ell} \Gamma_{i} M_{i j} \frac{\omega\tau_j }{\left[1+\left(\omega \tau_j\right)^2\right]} M_{j \ell}^{-1}\Gamma_\ell P_\ell^{\rm eq} \ . \label{eq: CN heat}
\end{align}
We assume the total energy stored in the system (averaged over one cycle) is dominated by the elastic energy~\eqref{eq: elastic energy}, so the inverse quality factor~\eqref{eq: quality factor definition} is
\begin{align}
	Q_{\rm CN}^{-1} = \frac{\beta N\gamma_{0}^2}{4\mcV C} \sum_{i,j,\ell} \Gamma_{i} M_{i j} \frac{\omega\tau_j }{\left[1+\left(\omega \tau_j\right)^2\right]} M_{j \ell}^{-1}\Gamma_\ell P_\ell^{\rm eq} \ . \label{eq: CN quality factor}
\end{align}
Equation \ref{eq: CN quality factor} is a central result of our work. The mechanical loss is decomposed into contributions from eigenmodes of the transition matrix $R$, each contributing to the overall mechanical loss. Since $R$ is a transition rate matrix of a connected network, it has exactly one zero eigenvalue with eigenvector corresponding to the equilibrium distribution. All other eigenvalues are negative (describing relaxation towards equilibrium) with eigenvectors that sum to zero (preserves normalization of probability). For an alternate derivation of this result, see Supplemental Material~\ref{app: Elastic Energy}.

The TLS model corresponds to a network where every pair of inherent structures are isolated. For this network there is no global equilibrium and each pair of connected inherent structures must be treated independently. In Supplemental Material~\ref{app: TLS derivation}, we calculate the quality factor for a network of two inherent structures, which when summed over all independent pairs yields a quality factor consistent with the popular TLS model that sums over individual TLS contributions ~\cite{damart2018}
\begin{align}
    Q_{\rm TLS}^{-1} = \frac{\beta\gamma_{0}^{{2}}}{4\mathcal{V} C} \sum_{i} \Gamma_i^2\operatorname{sech}^2\left(\frac{\beta\Delta_{i}}{2}\right)\frac{\omega \tau_{i}}{1+\omega^2 \tau_{i}^2} \ ,
	\label{eq: TLS Quality Factor}
\end{align}
with $\Delta_{i}$ the energy difference between the two inherent structures in TLS $i$ and $\Gamma_i$ their deformation potential, as well as the TLS relaxation time
\begin{align}
	\tau_{i}=\frac{e^{\beta V_{i}}}{k_{0}(1+e^{\beta\Delta_{i}})} \ .
\end{align}

In going from Eq.~\ref{eq: CN quality factor} to Eq.~\ref{eq: TLS Quality Factor} the $\sech^2$ term arises from the eigenvectors $M$ and the relative occupation probability of the two TLS through the Boltzmann factors in $P^{\rm eq}$.

The TLS mechanical loss can also be derived directly from the network description. Assuming every TLS $i$ has equal occupation probability and transitions independently, their transition rates $R^{\rm TLS}_{i}$ form $2\times 2$ matrices. The global transition rate matrix for $N$ independent Markov jump processes is the Kronecker sum of the constituent rate matrices $R = \oplus_{i = 1}^{N}R^{\rm TLS}_{i}$. The eigenvectors of the $2^N$ element matrix $R$ are the Kronecker product of the individual TLS eigenvectors with eigenvalues the corresponding sum of the individual TLS eigenvalues~\cite{Gardiner}. 

This can give rise to correlated eigenmodes involving multiple TLS eigenmodes. If each TLS also responds independently to mechanical perturbations, then the global coupling terms can be expressed as the sum of the individual TLS couplings. The sum of one-body functions never produces two-body (or higher) tensor components, and only the individual TLS eigenmodes contribute to mechanical loss. The special case of $N=2$ is treated in Supplemental Material~\ref{app: TLS derivation}. 

There are three assumptions that must be true for the TLS model to be valid: i) the equilibrium distribution $P^{\rm eq}$ is constant across different TLS, ii) the TLSs transition independently, and iii) the TLS couple independently to mechanical perturbations. These are strong assumptions given the disordered structure of amorphous solids and the potential for long range elastic couplings. We find that relaxing these assumptions dramatically changes the predicted mechanical loss.

\section{Model Systems}

There are two qualitatively novel features that arise from the network description of mechanical loss that we will highlight: i) the topology and connectivity of the network results in cycles and alternate pathways for exploring the energy landscape and ii) the energy levels of the inherent structures can be compared on a global scale.  Before we explore the mechanical loss in these systems, we study minimal models to clearly see the effect of the connectivity, topology, and energy level distribution. 

\subsection{Cycles and Connectivity: Four-State Networks}

\begin{figure}
	\includegraphics[width=\linewidth]{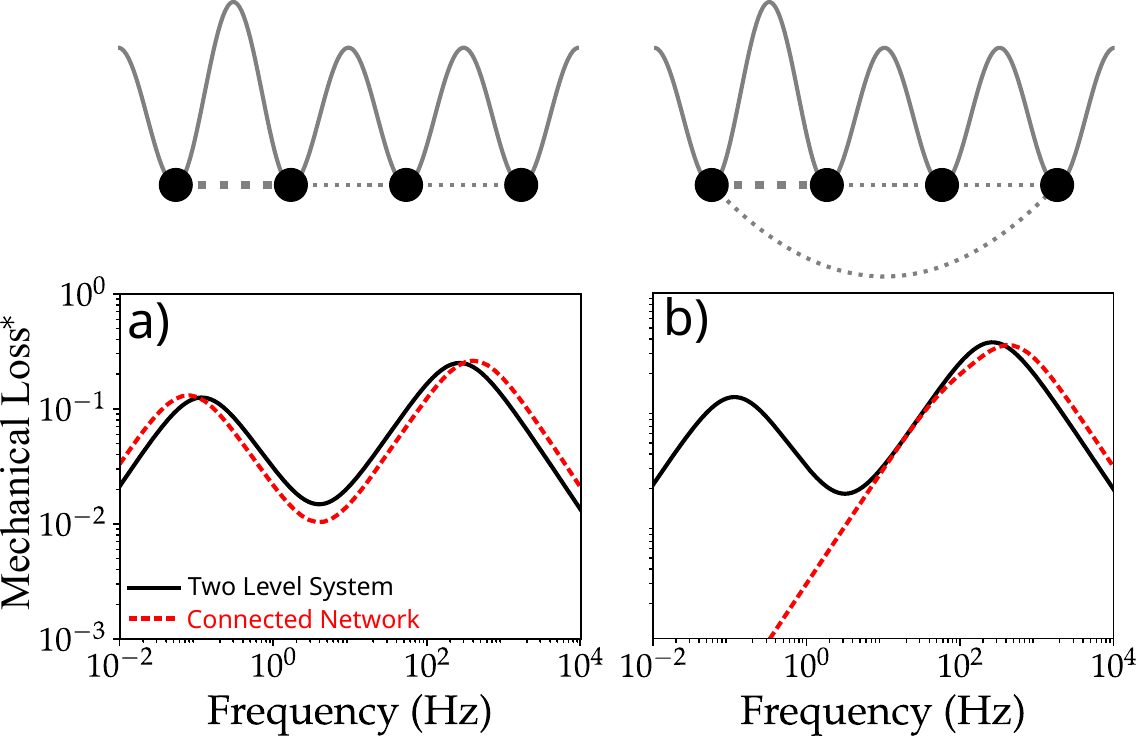}
	\caption{Scaled mechanical loss $Q^{-1}\mcV C/(\beta\gamma_{0}^2)$ as a function of frequency for a) a four-state chain and b) a four-state cycle. Two different curves are shown: the TLS approximation~\eqref{eq: TLS Quality Factor} (black solid line), and the connected network ~\eqref{eq: CN quality factor} (red dashed). In all cases the large energy barrier is $0.8$eV with the remaining barriers $0.6$eV.}
	\label{fig: chain_vs_Cycle_plot}
\end{figure}
Our atomistic samples of amorphous materials exhibit a significant fraction of basis cycles (Fig.~\ref{fig: TLS_Network}), with four-state cycles the most frequent. We therefore consider a minimal four-state network that consists of discrete states $i$ with energy minima $E_{i}$ and barriers $V_{ij}$ as depicted in Fig.~\ref{fig: chain_vs_Cycle_plot}. All states separated by a barrier are connected with transitions and dynamics given by eq.~\eqref{eq: ME General}. We estimate mechanical loss from eq.~\eqref{eq: CN quality factor} and eq.~\eqref{eq: TLS Quality Factor} and in all cases the temperature is $300K$. 

In Fig.~\ref{fig: chain_vs_Cycle_plot} we compare the mechanical loss predicted by the TLS and connected network models for a four-state chain (a) and a four-state cycle (b). The system has one barrier larger than the other equal barriers. This results in two timescales, one for each barrier height, which is reflected in the two peaks in dissipation in panel a). The large barrier causes the low frequency peak and the smaller barriers the higher frequency peak. Although the TLS approximation and connected network model predict the same qualitative behavior for the chain, dissipation differs significantly for the cycle. The low frequency peak is completely eliminated by the additional connection in the cycle. The added low barrier connection results in a high frequency pathway for the system to equilibrate, avoiding the large energy barrier.

To make concrete the connection between the topology of the energy landscape and the elimination of low frequency peaks, we analytically calculate and compare the eigenvalues of an equal energy minima three-state chain and a three-state cycle assuming one transition rate is small relative the remaining equal transition rates, i.e.~$x\equiv k_{12}/k_{0}\ll 1$:
\begin{align}
    R_{0}^{\rm chain} = k_{0}
\begin{bmatrix}
    -x& x & 0  \\
    x & -(x+1) & 1 \\
    0 & 1 & -1
\end{bmatrix} \ ,
\end{align}
and
\begin{align}
    R_{0}^{\rm cycle} = k_{0}
\begin{bmatrix}
    -(x+1)& x & 1  \\
    x & -(x+1) & 1 \\
    1 & 1 & -2
\end{bmatrix} \ .
\end{align}

In both cases the zero eigenvalue corresponds to the equilibrium distribution with equal probability in each state. For the chain, the remaining two eigenvalues are $\lambda_{2}^{\rm chain}\approx-xk_{0}$ and $\lambda_{3}^{\rm chain}\approx-2k_{0}$ with eigenvectors $[-2,1,1]$ and $[0,-1,1]$ respectively. The former results in a low-frequency peak in dissipation since $x$ is assumed to be small. For the cycle, the remaining two eigenvalues are $\lambda_{2}^{\rm cycle}\approx-3k_{0}$ and $\lambda_{3}^{\rm cycle}\approx-k_{0}$ with eigenvectors $[-1/2,-1/2,1]$ and $[-1,1,0]$ respectively. For the cycle the eigenvalues are approximately independent of the small relative transition rate $x$, effectively removing the low frequency peak observed in the chain. Similar calculations for a four-state chain yields eigenvalues $0,-4/3 x k_{0}, -k_{0}, -3k_{0}$ and for a four-state cycle $0,-2,-\sqrt{2}-2,\sqrt{2}-2$, showing an identical phenomenon.

Performing the same comparison between the four-state chain and cycle at fixed frequency and variable temperature would show the same qualitative trends. The low frequency peak becomes a high temperature peak in the mechanical loss-Temperature plane and the high frequency peak becomes a low temperature peak. The overall effect of connecting four states in a cycle can therefore be equally thought of as suppressing high temperature relaxation modes at fixed frequency. The next simplest structure formed by a four-state network would allow for connections between any of the four nodes. This results in a combination of three- and four-state cycles that behave similarly to the four-state cycle tested in this section.

\subsection{Cycles and Connectivity:  Barabási-Albert Networks}
The Barabási-Albert (BA) model is a widely used method for generating large, connected, and scale-free networks based on preferential attachment. The network is initialized with a small, fully connected set of $m_0$ nodes. New nodes are sequentially added to the network, and each new node forms $m$ edges with existing nodes. The probability $P^{\rm connect}_{i}$ that a new node connects to an existing node $i$ is proportional to the degree $d_i$ of that node:  
\begin{equation}
P^{\rm connect}_{i} = \frac{d_i}{\sum_{j} d_j},
\end{equation}
where the sum is over all existing nodes in the network. This preferential attachment mechanism ensures that nodes with higher degrees are more likely to gain new connections, resulting in the emergence of hubs and power-law degree distribution. By adjusting the parameters $m_0$ and $m$, the BA model can produce networks of varying sizes and densities while maintaining the scale-free property.

\begin{figure}
	\includegraphics[width=\linewidth]{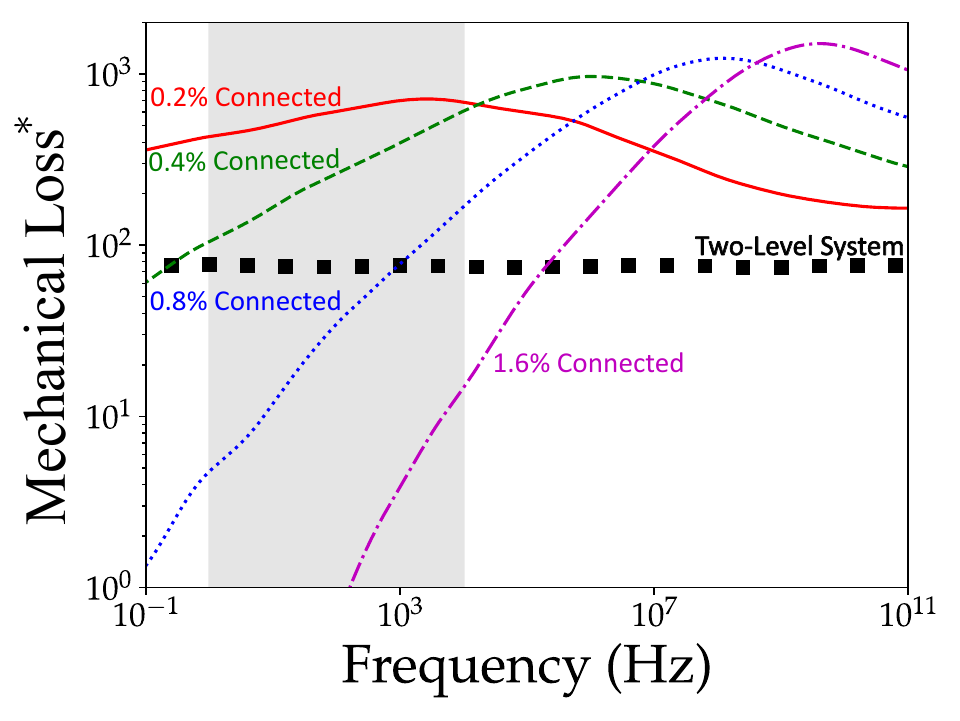}
	\caption{Average mechanical loss $Q^{-1}\mcV C/(\beta\gamma_{0}^2)$ from $1000$ node BA networks (curves) and an equivalent TLS model (black squares). Energies are drawn from a uniform energy distribution with barriers between states constrained to be larger than the energy minima. Results are averaged over 100 random networks. The connectivity of the network is increased by changing the number of connections between additional nodes when forming the network $m= 2,4,8,16$ for red solid, green dashed, blue dotted, and purple dash-dotted curves, respectively.}
	\label{fig: B_A_network_plot}
\end{figure}

In Fig.~\ref{fig: B_A_network_plot} we explore the effect of an increase in connectivity (number of connections/total number of possible connections)  on the predicted energy dissipation of random BA networks and compare with the simple TLS model. The energy levels of the states are drawn from a uniform distribution $E_{i} \in [0,0.5]$eV and the energy barriers from a uniform distribution satisfying $V_{ij} \in [\max(E_{i},E_{j})+0.01, \max(E_{i},E_{j})+1.01]$eV to ensure the energy barriers are always larger than the energy of the states it connects.

Since the energy barriers are drawn from a uniform distribution, the TLS model predicts a flat frequency spectrum. As connectivity increases from $0.2\%$ to $1.6\%$, low-frequency dissipation of the connected network decreases with a compensating increase in high-frequency dissipation. At the highest levels of connectivity, the dissipation is almost entirely eliminated from the frequency band relevant to gravitational wave detectors. The reduction in low-frequency dissipation is the result of the additional low energy barrier pathways between states allowing the network to avoid large energy barriers as discussed in the previous section. Just as with the four-state cycle, this can also be interpreted as reducing the high temperature mechanical loss at fixed frequency as the large barriers crossed by large thermal fluctuations are circumvented.

\subsection{Energy Distribution: Four-State Networks}

\begin{figure}
	\includegraphics[width=\linewidth]{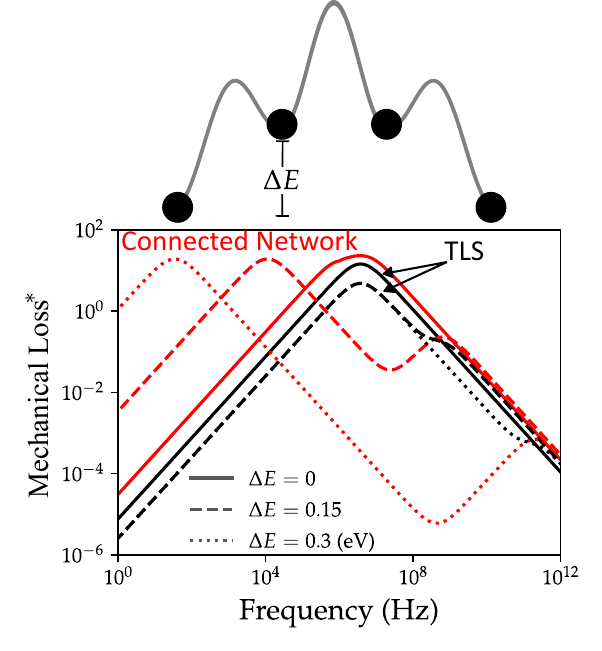}
	\caption{Mechanical loss $Q^{-1}\mcV C/(\beta\gamma_{0}^2)$ as a function of frequency. Two different colors (shades) of curves are shown: the TLS approximation~\eqref{eq: TLS Quality Factor} (black), and the connected network linear-response approximation~\eqref{eq: CN quality factor} (red). The energy of the center two minima is $0,0.15,0.3$ eV for solid, dashed, and dotted curves respectively. In all cases the energy barriers are $0.4$eV above the lower energy level connecting two states.}
	\label{fig: four-state mountain}
\end{figure}

In the previous section, we showed how the connectivity of the network can completely eliminate low frequency modes; however, if the distribution of energy minima is broad compared to the distribution of energy barriers the opposite effect occurs---the connected network results in low frequency modes not observed in the independent TLSs. As before, if the connectivity is increased far enough the low frequency modes are once again eliminated.

To illustrate this phenomenon we consider a four-state network with a ``mountain" landscape as depicted in Fig.~\ref{fig: four-state mountain}. Since the TLS model only compares pairs of energies, it has no sense of a global energy scale. Therefore, increasing the energy of the two minima in the center simply increases the asymmetry of two of the three TLS, decreasing dissipation without altering the frequency spectrum. 

In contrast, as the energy asymmetry between the center and outer minima is increased, a low frequency dissipation mode emerges in the connected network. For larger asymmetry, the frequency of the slow mode further decreases. As the center two minima are raised, the transition rate out towards the edges (down the landscape) increases while the transition rate towards the center (up the energy landscape) remains constant. The result is a slow relaxation mode between the two edges with very low probability of occupying the center (high energy) states.

More precisely, we can write the transition rate matrix as
\begin{align}
    R_{0}^{\rm mountain} = k_{0}
\begin{bmatrix}
    -1 & x & 0 & 0  \\
    1 & -(x+1) & 1 & 0 \\
    0 & 1 & -(x+1) & 1\\
    0 & 0 & x & -1
\end{bmatrix} \ ,
\end{align}
where we have assumed $k_{1\rightarrow2} = k_{2\rightarrow3} = k_{4\rightarrow3} = k_{0}$ and $k_{2\rightarrow1}=k_{3\rightarrow4} =k_{0}x$. As the energy level of the center two minima approaches the barrier height, the transition rate out of the state towards the edges becomes large, implying $x \gg 1$. In this case, the transition rate matrix has one slow mode with eigenvector $v_{\rm slow} \approx [-1,-1/x,1/x,1]$ and eigenvalue $\lambda_{\rm slow} \approx -2/x$. This corresponds to a slow mode that oscillates between the edges at a rate dependent on the rate out of the high energy states.

\subsection{Energy Distribution:  Barabási-Albert Networks}

\begin{figure}
	\includegraphics[width=\linewidth]{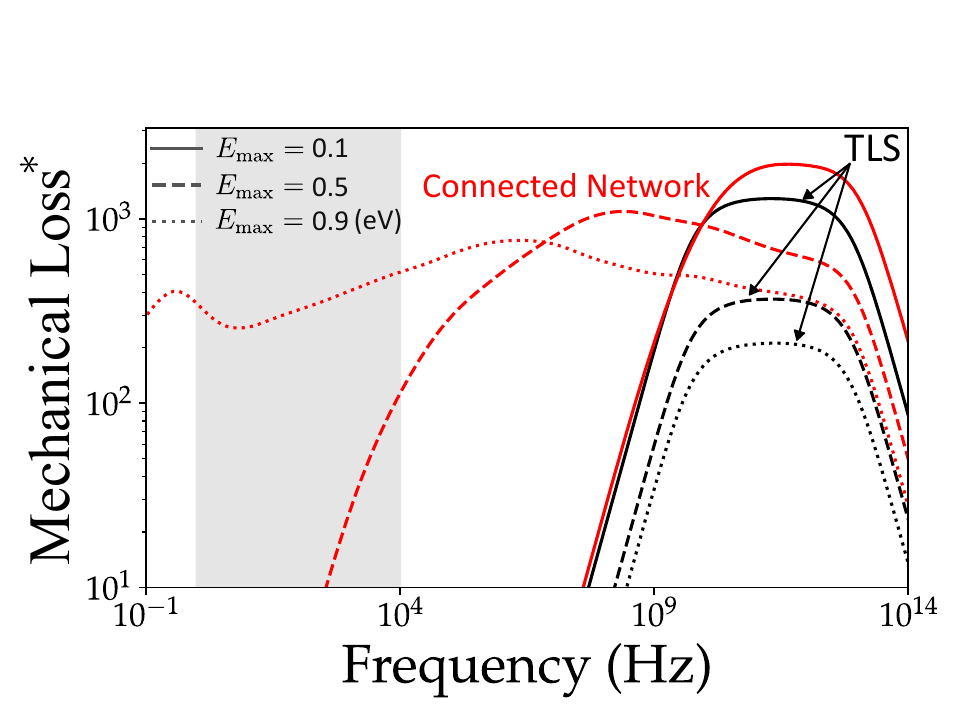}
	\caption{Average mechanical loss $Q^{-1}\mcV C/(\beta\gamma_{0}^2)$ from $1000$ node BA networks (red) and TLS (black). Here energies are drawn from a uniform  distribution between $0$ and $E_{\rm max}$, with $E_{\rm max} = 0.1,0.5,0.9$eV for solid, dashed, and dotted curves respectively. Barriers between states are constrained to be larger than the energy minima with a maximum of $0.2$eV. Results are averaged over 100 random networks. }
	\label{fig: B_A_network_broad_distribution}
\end{figure}

The low frequency modes resulting from the specific small-scale network considered in the previous section can be reproduced in larger random networks when the distribution of energy minima is broad compared to the energy barrier distribution. We generate BA networks with increasingly broad energy distributions in Fig.~\ref{fig: B_A_network_broad_distribution}. 

Increasing the width of the energy minima distribution decreases the dissipation predicted by the TLS model by increasing the asymmetry. In the connected networks, the broadening of the energy minima distribution also decreases the peak dissipation, since the increasing fraction of high energy states will have low equilibrium occupation probability and won't significantly contribute to the global dissipation. In addition, the broad energy minima distribution results in slow relaxation modes and low frequency peaks in dissipation just as we observed in the four-state ``mountain''.

\section{Amorphous silicon and titanium dioxide}
We now return to the atomistic models of a-Si and a-TiO$_{2}$ originally introduced in Fig.~\ref{fig: TLS_Network}. We consider 10 samples each containing 1,000 atoms for the a-Si samples and 750 atoms for the TiO$_{2}$ samples. We find the statistical properties of the networks to be robust between samples, but there are important differences between the two materials. In a-Si, the percentage of the network that is connected plateaus around $\sim 0.05\%$, while in a-TiO$_{2}$ the connectivity of the largest networks is $\sim0.14\%$. Although both materials exhibit a similarly broad distribution of energy minima, our a-Si models have no small barriers below $\sim 0.1$eV. Moreover, a-Si only features closed cycles with an even number of inherent structures, while samples of a-TiO$_{2}$ exhibit both even and odd state cycles.

Several generic properties of the mechanical loss can be inferred directly from Eq.~\eqref{eq: CN quality factor}. For low frequencies $\omega\tau \ll 1$ the mechanical loss scales as $\omega$, while for high frequencies $\omega\tau \gg 1$ it scales as $\omega^{-1}$. In the intermediate, regime there is a plateau region with peaks corresponding to $\omega\tau_{j} \sim 1$ for the relaxation modes that dissipate significant energy. To calculate mechanical loss, we use a typical value for the elastic modulus $C = 50$GPa~\cite{queen2015,fefferman2017,molina2021}, bare transition rate $k_{0}= 10^{13} {\rm s}^{-1}$~\cite{valiquette2003,levesque2022}, and estimate the deformation potential from differences in stress between connected inherent structures~\cite{damart2018} as outlined in Supplemental Material~\ref{app: simulation details}. We emphasize that these assumptions are made for calculations within both the TLS and connected network models, and changing the value of $k_{0}$ merely leads to an overall shift in the frequency $\omega$.

\begin{figure}
	\includegraphics[width=\linewidth]{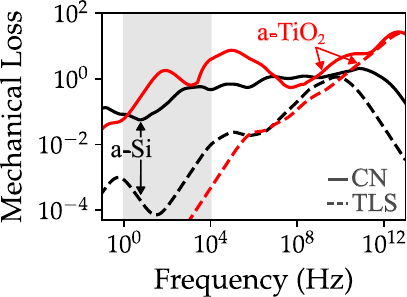}
    \caption{Mechanical loss $Q^{-1}$ of a-Si (black) and a-TiO$_{2}$ (red) for the connected network model (solid curves) and the TLS model (dashed curves) at 300K. Shaded in grey is the frequency band of interest for GWDs.}
	\label{fig: Combined_Heat}
\end{figure}

The inverse quality factor estimated from the connected network and TLS models found in Fig.~\ref{fig: Combined_Heat} have several qualitative and quantitative differences. It can be appreciated that in general, the TLS model underestimates the mechanical loss in comparison to the more comprehensive connected network models. For a-Si within the frequency range relevant for gravitational wave detection, the network model predicts non-monotonic behavior and a peak at $\sim 10^3$ Hz, while the TLS model predicts a peak at $\sim 1$ Hz and a mechanical loss that is overall smaller by two orders of magnitude. For a-TiO$_{2}$, the contrast is even more extreme as the TLS model would suggest vanishing mechanical loss in the same frequency range. This is due to an abundance of very small barriers that concentrate the mechanical loss in the high frequency regime. 

There are two competing effects that result in the significant differences between the TLS and connected network descriptions: i) the prevalence of cycles in the network (Fig.~\ref{fig: TLS_Network}) that are topologically distinct from a TLS and can allow the system to circumvent large barriers through alternate pathways reducing mechanical loss at low frequencies (Fig.~\ref{fig: chain_vs_Cycle_plot}), and ii) the broad distribution of energy minima (Fig.~\ref{fig: TLS_Network}) requires taking into account the global energy scale, which can lead to an increase in mechanical loss at low frequencies. 

Due to the relatively low connectivity of the networks formed by our amorphous samples (compared to those seen in Fig.~\ref{fig: B_A_network_plot}), the reduction in low frequency dissipation appears to be subdued. However, for our samples of a-TiO$_{2}$ we observe a significant increase in low frequency dissipation as a result of the broad distribution of energy minima resulting in slow relaxation modes (Fig.~\ref{fig: Combined_Heat}) analogous to the observed behavior of synthetic BA networks (Fig.~\ref{fig: B_A_network_broad_distribution}) and four-state cycles (Fig.~\ref{fig: four-state mountain}).

It is difficult to compute the mechanical loss~\eqref{eq: CN quality factor} at low temperatures since it requires diagonalizing large and nearly singular matrices. However, we can still understand the temperature behavior qualitatively by noting that low frequency modes are excited at high temperature (large barrier and large thermal fluctuations) and high frequency modes at low temperature (small barriers and small thermal fluctuations). This implies that at fixed frequency (e.g. $10^4$Hz) the TLS model predicts vanishing mechanical loss above $\sim 300$K for TiO$_{2}$ while the connected network has several high temperature relaxation modes (see Fig.~\ref{fig: Combined_Heat}).

\section{Discussion}
Motivated by the observation of a connected network of inherent structures coupled by thermally activated transitions in samples of amorphous solids, we have developed a rigorous generalization of the 50 year old TLS model of mechanical loss. Taking a nonequilibrium thermodynamic perspective, the dynamics of the discrete state network of inherent structures is described by a master equation (eq.~\eqref{eq: ME General}) for the time dependent probabilities, ultimately leading to an explicit expression for the energy dissipated in a single cycle of an acoustic oscillation (eq.~\eqref{eq: Heat}). The connected network model can describe hypercubic networks that would be formed by entirely independent TLS, but generalizes to TLS interacting via elastic or electrostatic forces and interacting multi-level systems in general.

The mechanical loss predicted in Fig.~\ref{fig: Combined_Heat} at $300$K and $1$kHz are of order $Q_{CN}^{-1} \sim10^{-1}$ and $\sim 10^{0}$ for a-Si and a-TiO$_{2}$ respectively. Experimental estimates for the mechanical loss of a-Si can vary significantly depending on film thickness and preparations conditions, ranging from $\sim10^{-3}$ for poorly annealed to $\sim 10^{-5}$ for well annealed samples~\cite{liu2014hydrogen,granata2020amorphous,molina2021}. This is a crucial factor to take into account when comparing our numerical predictions quantitatively to experiments, since the timescales accessible in MD simulations are significantly shorter than experimental protocols. 

The necessarily fast quench rates in MD simulations result in poorly annealed samples which we expect to increase the overall mechanical loss of our \emph{in silico} samples. The effect of aging and annealing has been the subject of several recent studies~\cite{khomenko2020,luckabauer2019,khadka2023,prasai2021}. Indeed, it has been shown that the density of TLS can be reduced by two orders of magnitude when comparing hyperquenched to ultrastable glass models~\cite{khomenko2020}. A promising direction for future study is therefore to explore how the effects of aging and annealing on mechanical loss manifests in the network structure of the materials. As we have shown, even the smallest networks consisting of three or four states can result in dramatic differences between TLS and connected network predictions, so we expect that even for these well annealed samples it will be important to correctly account for the connectivity between states.

Additionally, quantitative predictions are sensitive to the choice of empirical interatomic potentials used to model the energy landscape. One factor whose effect is easy to understand is the deformation potential $\gamma$, for which experiments estimates typically $\gamma \sim 1$eV, but some estimates can be as low as $\gamma \sim 7$meV~\cite{molina2021decoupling}. Since $\gamma$ enters quadratically in the calculation of the mechanical loss, the quantitative prediction depends strongly on the accuracy of this estimate, which in our simulations has a broad distribution (see appendix ~\ref{app: simulation details}). With the advancement of machine-learning potentials that are able to achieve \emph{ab initio} level accuracy, we are hopeful that this inaccuracy can be improved upon without sacrificing the accessible simulation timescales.

The lower mechanical loss predictions of the simple TLS model in certain frequency regimes in Fig.~\ref{fig: Combined_Heat}, and also found in previous studies~\cite{levesque2022,trinastic2016}, appear superficially to overlap more closely with the experimental range. In our opinion, this is not evidence that the TLS model is a more correct description of the system, quite the opposite: the connected network model simply relaxes an assumption of the TLS model that we find to be an inaccurate description of the underlying physical system. Since the TLS model fails both quantitatively and qualitatively under breaking the assumed independence of transition states, we need to re-examine earlier TLS model predictions.

Although the transition rates in our model and mechanical loss calculations are restricted to classical (stochastic) dynamics, the network itself makes no such assumption. It will be interesting to see if a similar model can be used to describe connected networks of tunneling transitions at low temperature, and what implications it would have for the tunneling TLS model~\cite{berthier2023}. This is an important question as low temperature TLS are believed to be a main cause of dielectric loss in quantum materials~\cite{muller2019,ku2005,cho2023}. Indeed, hints at the failing of the TLS model can be seen in the spectral diffusion and telegraphic dielectric loss channels in superconducting qubits, requiring extensions beyond the standard tunneling model~\cite{klimov2018fluctuations,muller2019,bejanin2021interacting}.

A major advantage of the network viewpoint is that it provides new properties to analyze in order to improve our physical understanding of internal friction in amorphous materials, potentially revealing new methods for the design of low mechanical loss coatings. Importantly, our model system studies have shown clearly that low frequency mechanical loss can be reduced by i) increasing the network connectivity and ii) reducing the width of the distribution of inherent state energy minima (relative to the distribution of energy barriers). The optimization of these material properties through material chemistry and processing pathways represent an important new challenge. Additionally, the robust nature of the statistical network properties opens up the possibility for the discovery of universal features of amorphous materials. One good candidate is the scale free degree distribution, which has been observed in similar materials~\cite{doye2002}.

\section*{Materials and Methods}\label{Sect: Materials and Methods}
We perform molecular dynamics simulations of 10 samples each with 1000 atoms of pure amorphous silicon and 750 atoms of titanium dioxide. a-Si is modeled using a Tersoff potential~\cite{tersoff1989}, while a Buckingham potential is used to describe a-TiO$_2$~\cite{matsui1991molecular}. Pure amorphous silicon can be produced by ion-implantation~\cite{laaziri1999}, and non-hydrogenated amorphous silicon is commonly made by sputter deposition~\cite{pawlewicz1978,bailey2015,fernandez2021}. An interesting extension would be to consider hydrogenated amorphous silicon as it can slightly reduce the mechanical loss and improve its optical properties~\cite{molina2023}. The LAMMPS code is used for all simulations~\cite{LAMMPS}. 

We prepare samples via melt-quench at a quench rate of $10^{11}K/s$ and find inherent structures by thermal search trajectories at $600K$ for $200$ps with a sampling frequency of $100$fs. We determine candidate TLS based on changes in the minimum energy and filter them based on participation ratio and maximum atomic displacement to remove unlikely candidates. Duplicate pairs of TLS are determined and removed based on a root-mean-squared atomic displacement criterion between structures less than $10^{-4}$\AA~for a-Si. For a-TiO$_{2}$ we use a displacement criterion of $10^{-3}$\AA~ and an additional energy difference criterion of $10^{-3}$eV. This procedure is common for amorphous sample preparation~\cite{deringer2018} and TLS calculations from MD simulation~\cite{levesque2022,damart2018}. 

To determine connected structures, we add an additional step in our analysis and use the same method for removing duplicate TLS to determine which inherent structures are identical. In this way, we connect TLS together (e.g. TLS A-B and B-C becomes A-B-C), thus forming connected networks of inherent structures as shown in Fig.~\ref{fig: TLS_Network}. For a-Si we run 100 search trajectories and find networks of $\sim 1000-3000$ inherent structures, and for TiO$_{2}$ we find considerably fewer and so we ran $200$ search trajectories and find networks with $\sim 10-1000$ inherent structures. Compared to previous studies~\cite{damart2018,levesque2022}, we report more connected inherent structures for three reasons: i) we allow for sequential transitions that diffuse the system further from the initial structure in state-space, ii) we do not filter based on asymmetries or barriers since all can contribute meaningfully to the mechanical loss in the connected network model, and iii) building on these previous studies, our thermal search trajectories were optimized for finding a large number of connected inherent structures. Further simulation details can be found in Supplemental Material~\ref{app: simulation details}.

We would like to emphasize that our search is by no means an exhaustive search of all the inherent structures. We have performed convergence tests for the fraction of the network that is connected and the fraction of cycles within the network and find them to be statistically robust  (Supplemental Material~\ref{app: convtests}).

\section*{Acknowledgments}
We thank Jess McIver, Jeff Young, and Ke Zou for many discussions of this work and a critical reading of the manuscript. This research was supported in part by the Canada First Research Excellence Fund, Quantum Materials and Future Technologies Program, the New Frontiers in Reserch Fund (Exploration stream), and the Discovery Grant program of the Natural Sciences and Engineering Research Council of Canada. Computational resources and services were provided by Advanced Research Computing at the University of British Columbia and the Digital Research Alliance of Canada (\url{alliancecan.ca}).

\bibliography{references}

@article{liu2014hydrogen,
  title = {Hydrogen-Free Amorphous Silicon with No Tunneling States},
  author = {Liu, Xiao and Queen, Daniel R. and Metcalf, Thomas H. and Karel, Julie E. and Hellman, Frances},
  journal = {Phys. Rev. Lett.},
  volume = {113},
  issue = {2},
  pages = {025503},
  numpages = {5},
  year = {2014},
  month = {Jul},
  publisher = {American Physical Society},
  doi = {10.1103/PhysRevLett.113.025503},
  url = {https://link.aps.org/doi/10.1103/PhysRevLett.113.025503}
}

@article{martinis2005decoherence,
  title = {Decoherence in Josephson Qubits from Dielectric Loss},
  author = {Martinis, John M. and Cooper, K. B. and McDermott, R. and Steffen, Matthias and Ansmann, Markus and Osborn, K. D. and Cicak, K. and Oh, Seongshik and Pappas, D. P. and Simmonds, R. W. and Yu, Clare C.},
  journal = {Phys. Rev. Lett.},
  volume = {95},
  issue = {21},
  pages = {210503},
  numpages = {4},
  year = {2005},
  month = {Nov},
  publisher = {American Physical Society},
  doi = {10.1103/PhysRevLett.95.210503},
  url = {https://link.aps.org/doi/10.1103/PhysRevLett.95.210503}
}

@article{granata2020amorphous,
  title={Amorphous optical coatings of present gravitational-wave interferometers},
  author={Granata, Massimo and Amato, Alex and Balzarini, Laurent and Canepa, Maurizio and Degallaix, J{\'e}r{\^o}me and Forest, Dani{\`e}le and Dolique, Vincent and Mereni, Lorenzo and Michel, Christophe and Pinard, Laurent and others},
  journal={Classical and Quantum Gravity},
  volume={37},
  number={9},
  pages={095004},
  year={2020},
  publisher={IOP Publishing}
}

@article{angelani1998connected,
  title={Connected network of minima as a model glass: Long time dynamics},
  author={Angelani, Luca and Parisi, Giorgio and Ruocco, Giancarlo and Viliani, Gabriele},
  journal={Physical Review Letters},
  volume={81},
  number={21},
  pages={4648},
  year={1998},
  publisher={APS}
}

@article{banerjee2012characterization,
  title={Characterization of the dynamics of glass-forming liquids from the properties of the potential energy landscape},
  author={Banerjee, Sumilan and Dasgupta, Chandan},
  journal={Physical Review E—Statistical, Nonlinear, and Soft Matter Physics},
  volume={85},
  number={2},
  pages={021501},
  year={2012},
  publisher={APS}
}

@book{Gardiner,
	author = {C.W. Gardiner},
 	title = {Handbook of Stochastic Methods for Physics, Chemistry and the Natural Sciences},
   	publisher = {Springer},
   	year = {1985},
   	edition = {2nd}
}

@article{Seifert2012,
  title={Stochastic thermodynamics, fluctuation theorems and molecular machines},
  author={Seifert, Udo},
  journal={Rep. Prog. Phys.},
  volume={75},
  number={12},
  pages={126001},
  year={2012},
  publisher={IOP Publishing}
}

@article{berthier2023,
	title={Modern computational studies of the glass transition},
	author={Berthier, Ludovic and Reichman, David R},
	journal={Nature Reviews Physics},
	volume={5},
	number={2},
	pages={102--116},
	year={2023},
	publisher={Nature Publishing Group UK London}
}

@article{damart2017,
	title={Theory of harmonic dissipation in disordered solids},
	author={Damart, T and Tanguy, A and Rodney, David},
	journal={Physical Review B},
	volume={95},
	number={5},
	pages={054203},
	year={2017},
	publisher={APS}
}

@article{damart2018,
	title={Atomistic study of two-level systems in amorphous silica},
	author={Damart, T and Rodney, D},
	journal={Physical Review B},
	volume={97},
	number={1},
	pages={014201},
	year={2018},
	publisher={APS}
}

@article{khomenko2020,
	title={Depletion of two-level systems in ultrastable computer-generated glasses},
	author={Khomenko, Dmytro and Scalliet, Camille and Berthier, Ludovic and Reichman, David R and Zamponi, Francesco},
	journal={Physical review letters},
	volume={124},
	number={22},
	pages={225901},
	year={2020},
	publisher={APS}
}

@article{luckabauer2019,
	title={Decreasing activation energy of fast relaxation processes in a metallic glass during aging},
	author={Luckabauer, Martin and Hayashi, Tomoki and Kato, Hidemi and Ichitsubo, Tetsu},
	journal={Physical Review B},
	volume={99},
	number={14},
	pages={140202},
	year={2019},
	publisher={APS}
}

@article{phillips1987,
	title={Two-level states in glasses},
	author={Phillips, William A},
	journal={Reports on Progress in Physics},
	volume={50},
	number={12},
	pages={1657},
	year={1987},
	publisher={IOP Publishing}
}

@article{deringer2018,
	title={Realistic atomistic structure of amorphous silicon from machine-learning-driven molecular dynamics},
	author={Deringer, Volker L and Bernstein, Noam and Bart{\'o}k, Albert P and Cliffe, Matthew J and Kerber, Rachel N and Marbella, Lauren E and Grey, Clare P and Elliott, Stephen R and Cs{\'a}nyi, G{\'a}bor},
	journal={The journal of physical chemistry letters},
	volume={9},
	number={11},
	pages={2879--2885},
	year={2018},
	publisher={ACS Publications}
}

@article{levesque2022,
	title={Internal mechanical dissipation mechanisms in amorphous silicon},
	author={L{\'e}vesque, Carl and Roorda, Sjoerd and Schiettekatte, Fran{\c{c}}ois and Mousseau, Normand},
	journal={Physical Review Materials},
	volume={6},
	number={12},
	pages={123604},
	year={2022},
	publisher={APS}
}

@article{doye2002,
	title={Network topology of a potential energy landscape: A static scale-free network},
	author={Doye, Jonathan PK},
	journal={Physical review letters},
	volume={88},
	number={23},
	pages={238701},
	year={2002},
	publisher={APS}
}

@article{khadka2023,
	title={Cryogenic mechanical loss of amorphous germania and titania-doped germania thin films},
	author={Khadka, S and Markosyan, A and Prasai, K and Dana, A and Yang, L and Tait, SC and Martin, IW and Menoni, CS and Fejer, MM and Bassiri, R},
	journal={Classical and Quantum Gravity},
	volume={40},
	number={20},
	pages={205002},
	year={2023},
	publisher={IOP Publishing}
}

@article{prasai2021,
	title={Annealing-Induced Changes in the Atomic Structure of Amorphous Silica, Germania, and Tantala Using Accelerated Molecular Dynamics},
	author={Prasai, Kiran and Bassiri, Riccardo and Cheng, Hai-Ping and Fejer, Martin M},
	journal={physica status solidi (b)},
	volume={258},
	number={9},
	pages={2000519},
	year={2021},
	publisher={Wiley Online Library}
}

@article{steinlechner2018,
	title={Development of mirror coatings for gravitational-wave detectors},
	author={Steinlechner, Jessica},
	journal={Philosophical Transactions of the Royal Society A: Mathematical, Physical and Engineering Sciences},
	volume={376},
	number={2120},
	pages={20170282},
	year={2018},
	publisher={The Royal Society Publishing}
}

@article{aspelmeyer2014,
	title={Cavity optomechanics},
	author={Aspelmeyer, Markus and Kippenberg, Tobias J and Marquardt, Florian},
	journal={Reviews of Modern Physics},
	volume={86},
	number={4},
	pages={1391},
	year={2014},
	publisher={APS}
}

@article{muller2019,
	title={Towards understanding two-level-systems in amorphous solids: insights from quantum circuits},
	author={M{\"u}ller, Clemens and Cole, Jared H and Lisenfeld, J{\"u}rgen},
	journal={Reports on Progress in Physics},
	volume={82},
	number={12},
	pages={124501},
	year={2019},
	publisher={IOP Publishing}
}

@article{ku2005,
	title={Decoherence of a Josephson qubit due to coupling to two-level systems},
	author={Ku, Li-Chung and Clare, C Yu},
	journal={Physical Review B},
	volume={72},
	number={2},
	pages={024526},
	year={2005},
	publisher={APS}
}

@article{cho2023,
	title={Simulating noise on a quantum processor: interactions between a qubit and resonant two-level system bath},
	author={Cho, Yujin and Jasrasaria, Dipti and Ray, Keith G and Tennant, Daniel M and Lordi, Vincenzo and DuBois, Jonathan L and Rosen, Yaniv J},
	journal={Quantum Science and Technology},
	volume={8},
	number={4},
	pages={045023},
	year={2023},
	publisher={IOP Publishing}
}

@article{tersoff1989,
	title={Modeling solid-state chemistry: Interatomic potentials for multicomponent systems},
	author={Tersoff, JJPRB},
	journal={Physical review B},
	volume={39},
	number={8},
	pages={5566},
	year={1989},
	publisher={APS}
}

@article{molina2021,
	title={Origin of mechanical and dielectric losses from two-level systems in amorphous silicon},
	author={Molina-Ruiz, M and Rosen, YJ and Jacks, HC and Abernathy, MR and Metcalf, TH and Liu, X and DuBois, JL and Hellman, F},
	journal={Physical Review Materials},
	volume={5},
	number={3},
	pages={035601},
	year={2021},
	publisher={APS}
}

@article{molina2021decoupling,
	title={Decoupling between propagating acoustic waves and two-level systems in hydrogenated amorphous silicon},
	author={Molina-Ruiz, Manel and Jacks, HC and Queen, DR and Metcalf, TH and Liu, Xiao and Hellman, Frances},
	journal={Physical Review B},
	volume={104},
	number={2},
	pages={024204},
	year={2021},
	publisher={APS}
}

@article{molina2023,
	title={Hydrogen-Induced Ultralow Optical Absorption and Mechanical Loss in Amorphous Silicon for Gravitational-Wave Detectors},
	author={Molina-Ruiz, Manel and Markosyan, Ashot and Bassiri, Riccardo and Fejer, MM and Abernathy, Matthew and Metcalf, TH and Liu, Xiao and Vajente, Gabriele and Ananyeva, Alena and Hellman, Frances},
	journal={Physical Review Letters},
	volume={131},
	number={25},
	pages={256902},
	year={2023},
	publisher={APS}
}

@article{LAMMPS,
	title={LAMMPS-a flexible simulation tool for particle-based materials modeling at the atomic, meso, and continuum scales},
	author={Thompson, Aidan P and Aktulga, H Metin and Berger, Richard and Bolintineanu, Dan S and Brown, W Michael and Crozier, Paul S and In't Veld, Pieter J and Kohlmeyer, Axel and Moore, Stan G and Nguyen, Trung Dac and others},
	journal={Computer Physics Communications},
	volume={271},
	pages={108171},
	year={2022},
	publisher={Elsevier}
}

@article{trinastic2016,
	title={Molecular dynamics modeling of mechanical loss in amorphous tantala and titania-doped tantala},
	author={Trinastic, Jonathan P and Hamdan, Rashid and Billman, Chris and Cheng, Hai-Ping},
	journal={Physical Review B},
	volume={93},
	number={1},
	pages={014105},
	year={2016},
	publisher={APS}
}

@article{queen2015,
	title={Two-level systems in evaporated amorphous silicon},
	author={Queen, DR and Liu, X and Karel, J and Jacks, HC and Metcalf, TH and Hellman, F},
	journal={Journal of Non-Crystalline Solids},
	volume={426},
	pages={19--24},
	year={2015},
	publisher={Elsevier}
}

@article{fefferman2017,
	title={Elastic measurements of amorphous silicon films at mk temperatures},
	author={Fefferman, Andrew and Maldonado, Ana and Collin, Eddy and Liu, Xiao and Metcalf, Tom and Jernigan, Glenn},
	journal={Journal of Low Temperature Physics},
	volume={187},
	pages={654--660},
	year={2017},
	publisher={Springer}
}

@article{valiquette2003,
  title={Energy landscape of relaxed amorphous silicon},
  author={Valiquette, Francis and Mousseau, Normand},
  journal={Phys. Rev. B},
  volume={68},
  number={12},
  pages={125209},
  year={2003},
  publisher={APS}
}

@article{laaziri1999,
  title={High resolution radial distribution function of pure amorphous silicon},
  author={Laaziri, Khalid and Kycia, S and Roorda, S and Chicoine, M and Robertson, JL and Wang, J and Moss, SC},
  journal={Phys. Rev. Lett.},
  volume={82},
  number={17},
  pages={3460},
  year={1999},
  publisher={APS}
}

@article{pawlewicz1978,
  title={Influence of deposition conditions on sputter-deposited amorphous silicon},
  author={Pawlewicz, WT},
  journal={Journal of Applied Physics},
  volume={49},
  number={11},
  pages={5595--5601},
  year={1978},
  publisher={American Institute of Physics}
}

@article{bailey2015,
  title={High rate amorphous and crystalline silicon formation by pulsed DC magnetron sputtering deposition for photovoltaics},
  author={Bailey, Louise R and Proudfoot, Gary and Mackenzie, Brodie and Andersen, Niels and Karlsson, Arne and Ulyashin, Alexander},
  journal={physica status solidi (a)},
  volume={212},
  number={1},
  pages={42--46},
  year={2015},
  publisher={Wiley Online Library}
}

@article{fernandez2021,
  title={Sputtered non-hydrogenated amorphous silicon as alternative absorber for silicon photovoltaic technology},
  author={Fern{\'a}ndez, Susana and Gand{\'\i}a, J Javier and Saugar, El{\'\i}as and G{\'o}mez-Mancebo, M{\textordfeminine} Bel{\'e}n and Canteli, David and Molpeceres, Carlos},
  journal={Materials},
  volume={14},
  number={21},
  pages={6550},
  year={2021},
  publisher={MDPI}
}

@article{vajente2021low,
  title={Low mechanical loss TiO 2: GeO 2 coatings for reduced thermal noise in gravitational wave interferometers},
  author={Vajente, Gabriele and Yang, Le and Davenport, Aaron and Fazio, Mariana and Ananyeva, Alena and Zhang, Liyuan and Billingsley, Garilynn and Prasai, Kiran and Markosyan, Ashot and Bassiri, Riccardo and others},
  journal={Physical Review Letters},
  volume={127},
  number={7},
  pages={071101},
  year={2021},
  publisher={APS}
}

@article{raza2015,
  title={Computer simulations of glasses: the potential energy landscape},
  author={Raza, Zamaan and Alling, Bj{\"o}rn and Abrikosov, Igor A},
  journal={Journal of Physics: Condensed Matter},
  volume={27},
  number={29},
  pages={293201},
  year={2015},
  publisher={IOP Publishing}
}

@article{murray2015,
  title={Ion-beam sputtered amorphous silicon films for cryogenic precision measurement systems},
  author={Murray, Peter G and Martin, Iain W and Craig, Kieran and Hough, James and Robie, Raymond and Rowan, Sheila and Abernathy, Matt R and Pershing, Teal and Penn, Steven},
  journal={Physical Review D},
  volume={92},
  number={6},
  pages={062001},
  year={2015},
  publisher={APS}
}

@article{matsui1991molecular,
  title={Molecular dynamics simulation of the structural and physical properties of the four polymorphs of TiO2},
  author={Matsui, Masanori and Akaogi, Masaki},
  journal={Molecular Simulation},
  volume={6},
  number={4-6},
  pages={239--244},
  year={1991},
  publisher={Taylor \& Francis}
}

@article{bejanin2021interacting,
  title={Interacting defects generate stochastic fluctuations in superconducting qubits},
  author={B{\'e}janin, JH and Earnest, CT and Sharafeldin, AS and Mariantoni, M},
  journal={Physical Review B},
  volume={104},
  number={9},
  pages={094106},
  year={2021},
  publisher={APS}
}

@article{klimov2018fluctuations,
  title={Fluctuations of energy-relaxation times in superconducting qubits},
  author={Klimov, Paul V and Kelly, Julian and Chen, Zijun and Neeley, Matthew and Megrant, Anthony and Burkett, Brian and Barends, Rami and Arya, Kunal and Chiaro, Ben and Chen, Yu and others},
  journal={Physical review letters},
  volume={121},
  number={9},
  pages={090502},
  year={2018},
  publisher={APS}
}

\setcounter{section}{0}

\onecolumngrid
\clearpage
\begin{center}
	\textbf{\large Supplemental Material for ``Connected Network Model for the Mechanical Loss of Amorphous Materials''}
\end{center}
\setcounter{equation}{0}
\setcounter{figure}{0}
\setcounter{table}{0}
\setcounter{page}{1}
\makeatletter
\renewcommand{\theequation}{S\arabic{equation}}
\renewcommand{\thefigure}{S\arabic{figure}}

\section{Elastic Energy}\label{app: Elastic Energy}

In this section, we present an alternate derivation of the quality factor~\eqref{eq: CN quality factor} based on the phase-lag of the average energy of the system. As discussed in the main text~\eqref{eq: elastic energy}, the total energy of the system is the sum of the elastic work and the average energy of the network
\begin{align}
	\mcU_{\rm tot} = \mcW_{\rm elastic} + \mcU_{\rm CN} \ .
\end{align}
Substituting in the linear elastic work Eq.~\eqref{eq: elastic energy}, energy of the connected network Eq.~\eqref{eq: Time Dependent Energy}, and assuming the approximation of Eq.~\eqref{eq: ME approx} we have
\begin{align}
	\mcU_{\rm tot} =  \mcV C \epsilon_0^2\sin^2(\omega t) + N\left[\boldsymbol{E}(0) + \frac{\epsilon_0 \gamma_0}{2}\boldsymbol{\Gamma}\sin(\omega t)\right] \cdot\left[\boldsymbol{P}^{\rm eq}-\frac{\beta\epsilon_0\gamma_0}{2}\boldsymbol{A} \sin (\omega t)-\frac{\beta\epsilon_0\gamma_0}{2}\boldsymbol{B} \cos (\omega t)\right] \ .
\end{align}
Expanding and collecting terms we have
\begin{align}
	\mcU_{\rm tot} = N\boldsymbol{E}(0)\cdot\boldsymbol{P}^{\rm eq} + \mcU_{\rm tot}^{(1)}\sin(\omega t-\theta) + \mcU_{\rm tot}^{(2)}\sin(\omega t-\phi)\sin(\omega t) \ ,
\end{align}
for 
\begin{align}
	\mcU_{\rm tot}^{(1)} &\equiv \frac{N\epsilon_0\gamma_0}{2}\sqrt{(\boldsymbol{\Gamma}\cdot \boldsymbol{P}^{\rm eq}-\beta\boldsymbol{E}(0)\cdot\boldsymbol{A})^2 + (\beta\boldsymbol{E}(0)\cdot\boldsymbol{B})^2} \\ 
	\theta &\equiv \tan^{-1}\left[\frac{\beta\boldsymbol{E}(0)\cdot\boldsymbol{B}}{\boldsymbol{\Gamma}\cdot \boldsymbol{P}^{\rm eq}-\beta\boldsymbol{E}(0)\cdot\boldsymbol{A}}\right] \\
	\mcU_{\rm tot}^{(2)} &\equiv\frac{N\epsilon_0^2\gamma_{0}^2}{4} \sqrt{(\mcV C -\beta\boldsymbol{\Gamma}\cdot\boldsymbol{A})^2 + (\beta\boldsymbol{\Gamma}\cdot\boldsymbol{B})^2} \\ 
	\phi &\equiv \tan^{-1}\left[\frac{\beta N\gamma_0^2\boldsymbol{\Gamma}\cdot\boldsymbol{B}}{4\mcV C -\beta N\gamma_0^2\boldsymbol{\Gamma}\cdot\boldsymbol{A}}\right] \ .
\end{align}
The constant and linear terms average out to zero over one cycle and have no contribution to the overall mechanical loss. Substituting Eq.~\eqref{eq: Probability approx} and assuming $\mcV C \gg \beta N\gamma_0^2\boldsymbol{\Gamma}\cdot \boldsymbol{A}/4$ we find
\begin{align}
	\tan\phi = Q^{-1} = \frac{\beta N\gamma_0^2}{\mcV C} \sum_{i,j,\ell} \Gamma_{i} M_{i j} \frac{\omega\tau_j }{\left[1+\left(\omega \tau_j\right)^2\right]} M_{j \ell}^{-1}\Gamma_\ell P_\ell^{\rm eq} \ . 
\end{align}
The inverse quality factor can be expressed in terms of the phase lag of the system $\phi$ relative to the frequency of the oscillation $\omega$: the mechanical loss results from the nonequilibrium, out of phase response of the system. The assumption $\mcV C \gg \beta\gamma_{0}\boldsymbol{\Gamma}\cdot \boldsymbol{A}/4$ is identical to the one made to arrive at equation (1) in Ref.~\cite{damart2018} as shown in their equation (B6).

\section{TLS derivation}\label{app: TLS derivation}

In this section we explicitly derive the mechanical loss for a TLS based on Eq.~\eqref{eq: CN quality factor} for a two-state system. The transition rate matrix, Eq.~\eqref{eq: transition matrix}, for a TLS consisting of state $1$ and $2$ is
\begin{align}
	R = k_{0}
	\begin{bmatrix}
		-e^{-\beta (V-E_{1})}      & e^{-\beta (V-E_{2})} \\
		e^{-\beta (V-E_{1})}      & -e^{-\beta (V-E_{2})} 
	\end{bmatrix} \ .
\end{align}
Defining the energy asymmetry $\Delta = E_2 - E_1$ this simplifies to
\begin{align}
	R = k_{0}e^{-\beta V}
	\begin{bmatrix}
		-1      & e^{\beta \Delta} \\
		1     & -e^{\beta \Delta} 
	\end{bmatrix}  \ ,
\end{align}
where, without loss of generality, we have set $E_{1} = 0$. This transition rate matrix has eigenvalues $\lambda_{1} = 0$ and $\lambda_{2} = -k_{0}e^{-\beta V}(1 + e^{\beta\Delta})$ with corresponding eigenvectors $\boldsymbol{v}^{(1)} = \left[1,e^{-\beta\Delta}\right]$ and $\boldsymbol{v}^{(2)} = \left[-1,1\right]$. The zero eigenvalue corresponds to the equilibrium distribution, so $\boldsymbol{P}^{\rm eq} = \boldsymbol{v}^{(1)}/\sum_{i}{v}^{(1)}_i$.

The eigenvector matrix and its inverse are
\begin{align}
	M =
	\begin{bmatrix}
		1      & -1 \\
		e^{-\beta\Delta}      & 1
	\end{bmatrix}  
\end{align}
and
\begin{align}
	M^{-1} =\frac{1}{1+e^{-\beta\Delta}}
	\begin{bmatrix}
		1      & 1 \\
		-e^{-\beta\Delta}      & 1
	\end{bmatrix}  \ .
\end{align}
setting $\boldsymbol{\Gamma} = \left[1,-1\right]$ (structures oscillate in opposite direction), substituting into Eq.~\eqref{eq: CN quality factor}, defining $\tau = -1/\lambda_{2}$ and summing over all TLS we arrive at Eq.~\eqref{eq: TLS Quality Factor}. Note that the zero eigenvalue mode has no contribution to the mechanical loss.
\begin{figure}
	\includegraphics[width=0.5\linewidth]{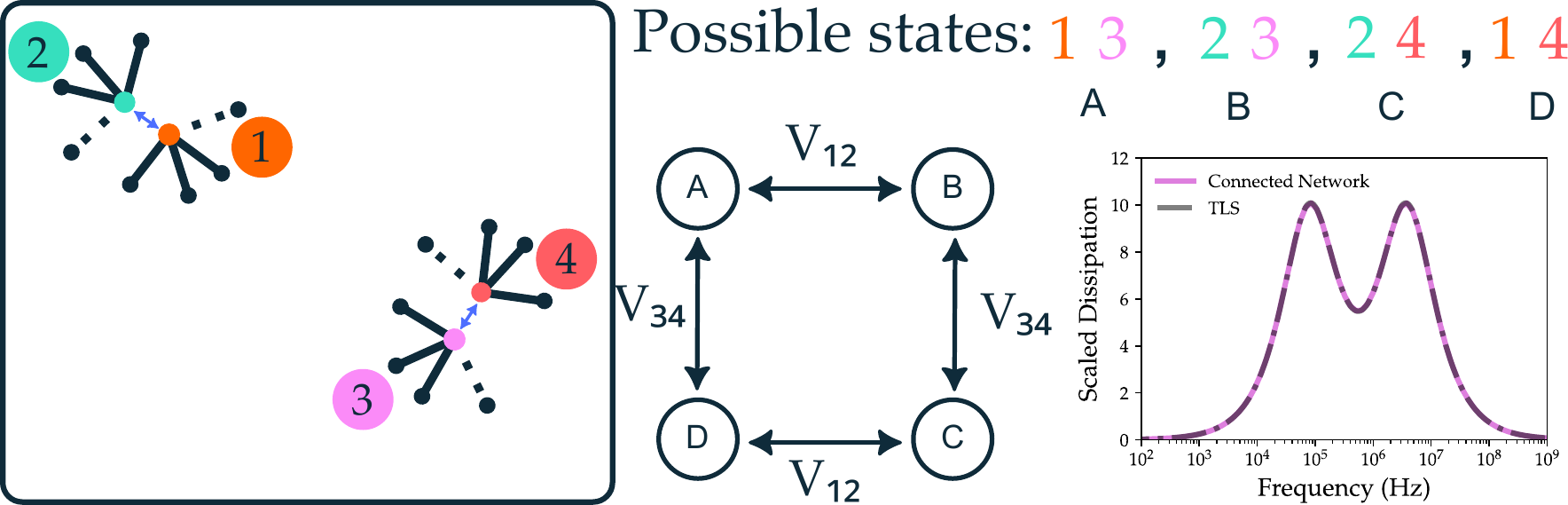}
    \caption{Idealized example of two isolated bond-hopping transitions (left) resulting in four inherent structures A-D. These inherent structures are connected in a symmetric four-state loop, which when driven by $\boldsymbol{\Gamma} = \tfrac{1}{2}(\gamma_{\alpha}\boldsymbol{v}_{\alpha}+\gamma_{\beta}\boldsymbol{v}_{\beta})$ produces identical mechanical loss (scaled dissipation $Q^{-1}\mcV C/\gamma_{0}^2$) with the TLS (black) or connected network (purple) models.}
	\label{fig: TLS Comparison}
\end{figure}

Independent TLS transitions can result in many distinct inherent structures. For example, two defects with corresponding transitions separated by a very large spatial distance forms two independent TLSs. Labeling the allowed transitions 1$\leftrightarrow$2 (TLS $\alpha$) and 3$\leftrightarrow$4 (TLS $\beta$), there are 4 allowed states in the global potential energy landscape (inherent structures): (1,3) A, (1,4) B, (2,4) C, (2,3) D. The allowed transitions form a closed loop ...-A-B-C-D-A-... (Fig.~\ref{fig: TLS Comparison}). The allowed relaxation modes of such a network recover the two independent relaxation modes of the TLSs $\lambda_{\alpha}$ and $\lambda_{\beta}$, with an additional correlated relaxation mode $\lambda_{3}$. 
\begin{align}
    v_{0} &= [1,1,1,1] \ , \ \lambda_{0} = 0  \ ,\\
    v_{\alpha} &= [1,-1,-1,1] \ , \ \lambda_{\alpha} \ ,\\
    v_{\beta} &= [1,1,-1,-1] \ , \ \lambda_{\beta} \ ,\\
    v_{\gamma} &= [1,-1,1,-1] \ , \ \lambda_{\gamma} = \lambda_{\alpha}+\lambda_{\beta} \ .
\end{align}

If the two transitions are independent, then they couple independently to incident mechanical waves and we have 
\begin{align}
\Gamma_{A} &= \frac{1}{2}\left(\Gamma_{1} + \Gamma_{3} \right), \\ 
\Gamma_{B} &= \frac{1}{2}\left(\Gamma_{1} + \Gamma_{4} \right), \\
\Gamma_{C} &= \frac{1}{2}\left(\Gamma_{2} + \Gamma_{4} \right), \\
\Gamma_{D} &= \frac{1}{2}\left(\Gamma_{2} + \Gamma_{3} \right) \ .
\end{align}
Since $\boldsymbol{\Gamma}\cdot \boldsymbol{v}_{\gamma} = \Gamma_{A}-\Gamma_{B}+\Gamma_{C}-\Gamma_{D} = 0$, the third correlated relaxation mode does not contribute to the mechanical loss and we recover the TLS model description. This coupling is estimated directly from simulations for a-Si and a-TiO$_{2}$ calculations.

\section{Simulation details}\label{app: simulation details}
In this section we provide additional simulation details. By injecting random initial velocities, ten amorphous samples of 1000 atoms each of a-Si and 750 atoms each of TiO$_{2}$ are prepared by rapidly melting diamond silicon and rutile TiO$_{2}$ from $100$K to $4000$K in $200$ps, equilibrating at $4000$K for $200$ps, and subsequently cooling to $300$K at a rate of $10^{11}$ K/s in the isobaric ensemble (target pressure P=0). Final amorphous configurations are found by energy minimization of the melt-quenched structure at constant volume. 

For a-Si we have an average simulation box length of $27.43 \pm0.01$\AA, the density of our samples is $2.261{\rm g/cm^3} \pm 0.001$. Using a cut off of $2.9$\AA~the silicon atoms in the $10$ samples have average coordination (with standard error) $c_{3} = 0.42 \pm 0.08 \%$, $c_4 = 95.8 \pm 0.2\%$, $c_5 = 3.8 \pm 0.2\%$, and $c_6 = 0.03 \pm 0.02\%$, where $c_{i}$ is the percentage of the sample with coordination $i$. This corresponds to $\sim 4\%$ defects in our samples. For a-TiO$_{2}$ we have a simulation box of lengths $22.97\times22.97\times14.79\pm0.01$\AA, for an average density of $4.247{\rm g/cm^3}\pm 0.001$.

\begin{figure}
	\includegraphics[width=0.5\linewidth]{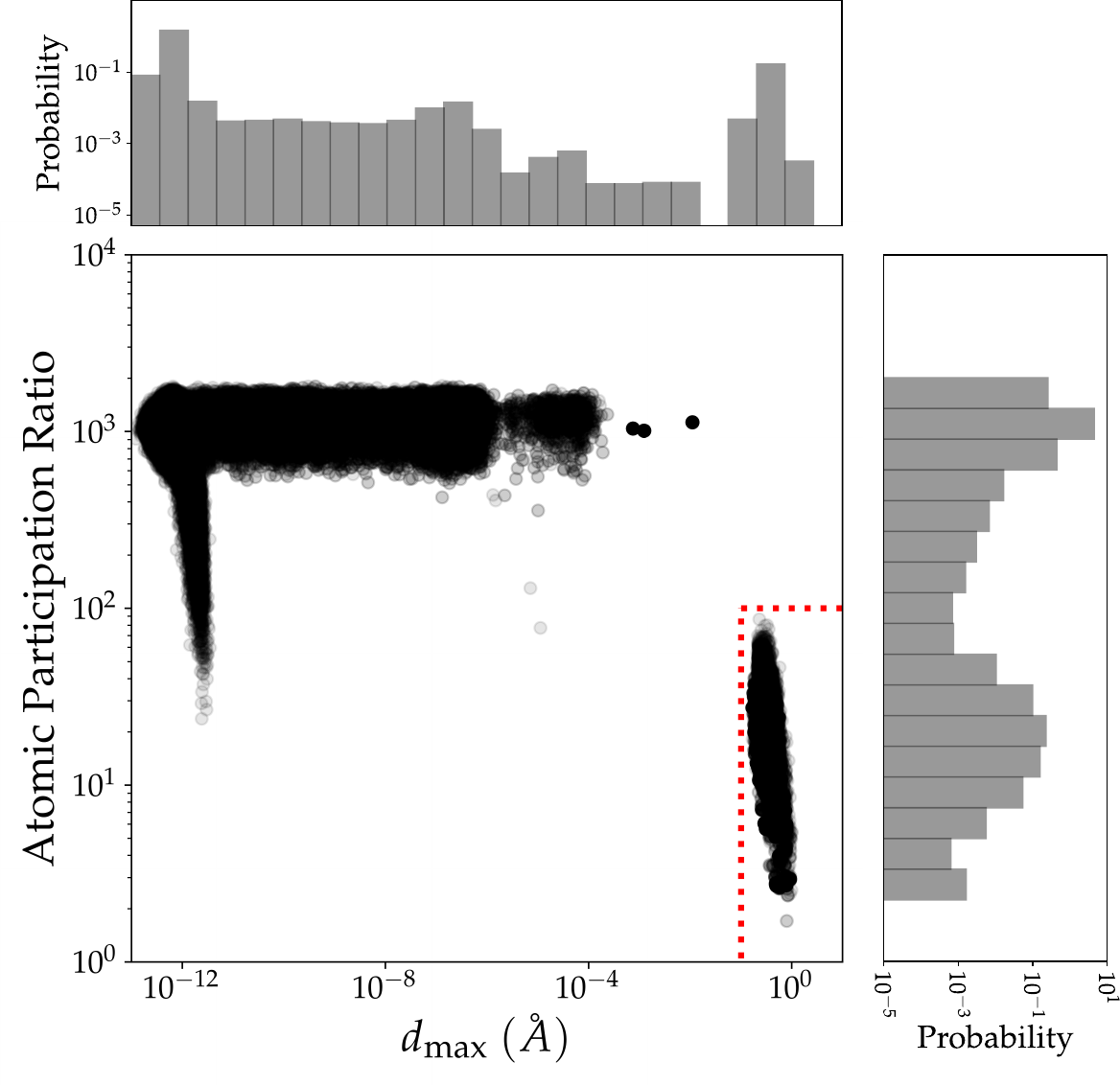}
	\caption{Distribution of participation ratio and maximum atomic displacement $d_{\rm max}$ between candidate connected inherent structures of a-Si. The region enclosed by dotted red lines ($PR <100$ and $d_{\rm max} > 0.1$\AA) are selected as likely candidates for being connected pairs.}
	\label{fig: PR_vs_dmax}
\end{figure}

Once the samples have been prepared, we perform $100$ random thermal searches for a-Si and 200 for a-TiO$_{2}$ per sample at $600$K for $200$ps. Every $0.1$ps we save the structure of the system and the $2000$ structures per search ($200,000$ per sample a-Si,$400,000$ per sample a-TiO$_{2}$) are quenched to $0$K providing an inherent structure of the system. Sequentially visited structures are considered as candidate connected pairs, and the atomic participation ratio (number of atoms involved in the transition) and maximum atomic displacement between these pairs is calculated and used to filter out unlikely candidates as shown in Fig.~\ref{fig: PR_vs_dmax}.

\begin{figure}
	\includegraphics[width=0.5\linewidth]{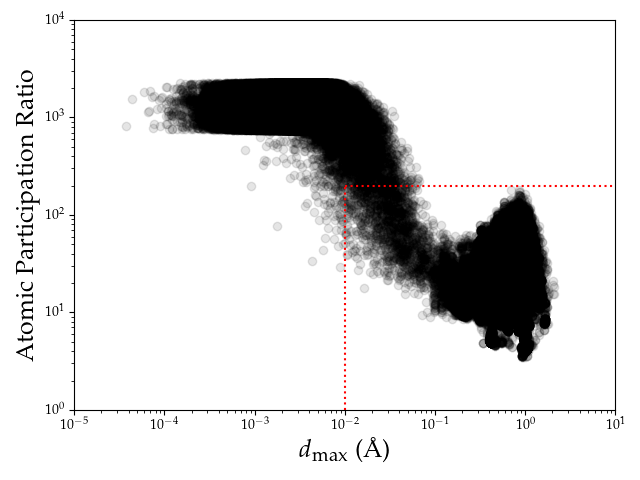}
	\caption{Distribution of participation ratio and maximum atomic displacement $d_{\rm max}$ between candidate connected inherent structures of a-TiO$_{2}$. The region enclosed by dotted red lines ($PR <200$ and $d_{\rm max} > 0.01$\AA) are selected as likely candidates for being connected pairs.}
	\label{fig: PR_vs_dmax Ti}
\end{figure}

In our a-Si data we observe two distinct regions in participation ratio-$d_{\rm max}$ space: pairs of structures with large participation ratio $\sim 10^{3}$ and small maximum atomic displacement and low participation ratio $\lesssim 100$ and comparatively large $d_{\rm max} \gtrsim 0.1$. The former corresponds to all the atoms moving a very small distance and is likely the result of noise, while the latter involves relatively few atoms moving a larger distance. 

\begin{figure}
	\includegraphics[width=0.5\linewidth]{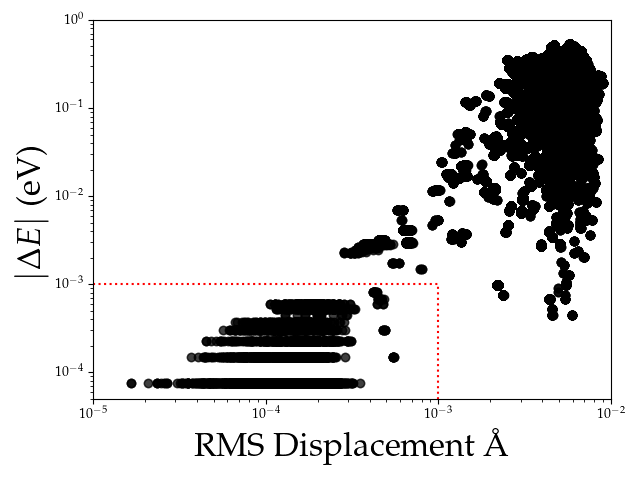}
	\caption{Distribution of energy difference and total root-mean-squared (RMS) displacement between inherent structures of a-TiO$_{2}$. The region enclosed by dotted red lines ($|\Delta E| < 10^{-3}$eV and RMS displacement $< 10^{-3}$\AA) are considered the same atomic structure.}
	\label{fig: Energy_vs_RMSD Ti}
\end{figure}

For a-Si, we consider all the candidates with participation ratio $<100$ and $d_{\rm max} > 0.1$\AA. For a-TiO${2}$, we do not observe as clear a cut off, so we use a generous cut off of participation ratio $<200$ and $d_{\rm max} > 0.01$\AA~ (Fig.~\ref{fig: PR_vs_dmax Ti}). We then remove all duplicate pairs of atomic structures based on total root-mean squared atomic displacement of $10^{-4}$\AA~ for a-Si. We found that a-TiO$_{2}$ required a root-mean squared atomic displacement of $10^{-3}$\AA~ with an additional energy difference criteria of $\Delta E < 10^{-3}$eV to identify duplicate structures (Fig.~\ref{fig: Energy_vs_RMSD Ti}). 

From this filtered list, we perform nudged elastic band calculations with $32$ intermediate structures to determine the transition path and barrier between the two states. If we find only a single maximum between the two states and the energy of that maximum is larger than the energy of both structures then we accept the pair as connected structures. The full distribution of barriers and asymmetries between connected inherent structures are shown in the main text.

To form the connected network, we calculate the root-mean squared total atomic displacement between all remaining inherent structures. If it is less than $10^{-4}$\AA~ for a-Si or $10^{-3}$\AA~ with an additional energy difference criteria of $\Delta E < 10^{-3}$eV for TiO$_{2}$, then we assume they are the same inherent structure. This connects TLS together since distinct pairs of inherent structures (a TLS) often share one inherent structure. For example, the two TLS A-B and C-D would merge to form A-B-C if we determined state B and C were the same inherent structure.

The longitudinal component of the deformation potential is estimated from differences in stress $\Delta\sigma^{(ij)}$ between inherent structures $i$ and $j$ as suggested in ref.~\cite{damart2018}:
\begin{align}
    (\gamma^{\rm L}_{0})^2\Gamma^{\rm L}_{i}\Gamma^{\rm L}_{j} = &\frac{\mcV^2}{5}\left[(\Delta\sigma^{(ij)}_{xx})^2 +(\Delta\sigma^{(ij)}_{yy})^2 +(\Delta\sigma^{(ij)}_{zz})^2\right] + \frac{2\mcV^2}{15}\left[\Delta\sigma^{(ij)}_{xx}\Delta\sigma^{(ij)}_{yy} +\Delta\sigma^{(ij)}_{xx}\Delta\sigma^{(ij)}_{zz} +\Delta\sigma^{(ij)}_{yy}\Delta\sigma^{(ij)}_{zz}\right] \nonumber\\
    &+ \frac{4\mcV^2}{15}\left[(\Delta\sigma^{(ij)}_{xy})^2 +(\Delta\sigma^{(ij)}_{xz})^2  +(\Delta\sigma^{(ij)}_{yz})^2 \right] \ .
    \label{eq: Longitudinal deformation potential}
\end{align}
Similar calculation yields the transverse component
\begin{align}
    (\gamma^{\rm T}_{0})^2\Gamma^{\rm T}_{i}\Gamma^{\rm T}_{j} = &\frac{\mcV^2}{15}\left[(\Delta\sigma^{(ij)}_{xx})^2 +(\Delta\sigma^{(ij)}_{yy})^2 +(\Delta\sigma^{(ij)}_{zz})^2\right] - \frac{\mcV^2}{15}\left[\Delta\sigma^{(ij)}_{xx}\Delta\sigma^{(ij)}_{yy} +\Delta\sigma^{(ij)}_{xx}\Delta\sigma^{(ij)}_{zz} +\Delta\sigma^{(ij)}_{yy}\Delta\sigma^{(ij)}_{zz}\right] \nonumber\\
    &+ \frac{3\mcV^2}{15}\left[(\Delta\sigma^{(ij)}_{xy})^2 +(\Delta\sigma^{(ij)}_{xz})^2  +(\Delta\sigma^{(ij)}_{yz})^2 \right] \ .
\end{align}

\begin{figure}
	\includegraphics[width=0.5\linewidth]{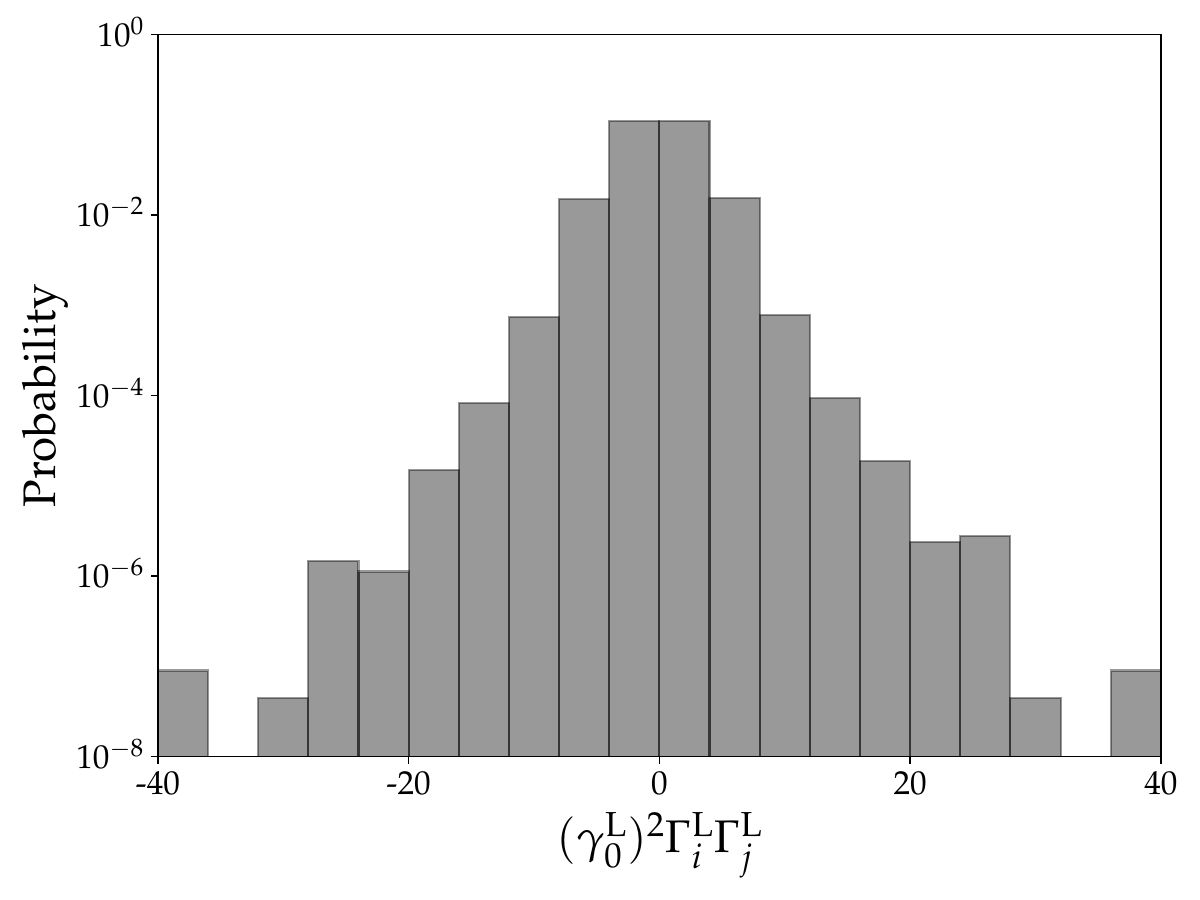}
	\caption{Histogram of the longitudinal squared deformation potential~\eqref{eq: Longitudinal deformation potential} in eV$^2$ for a single sample of amorphous silicon.}
	\label{fig: Gamma_square_distribution}
\end{figure}

\begin{figure}
	\includegraphics[width=0.5\linewidth]{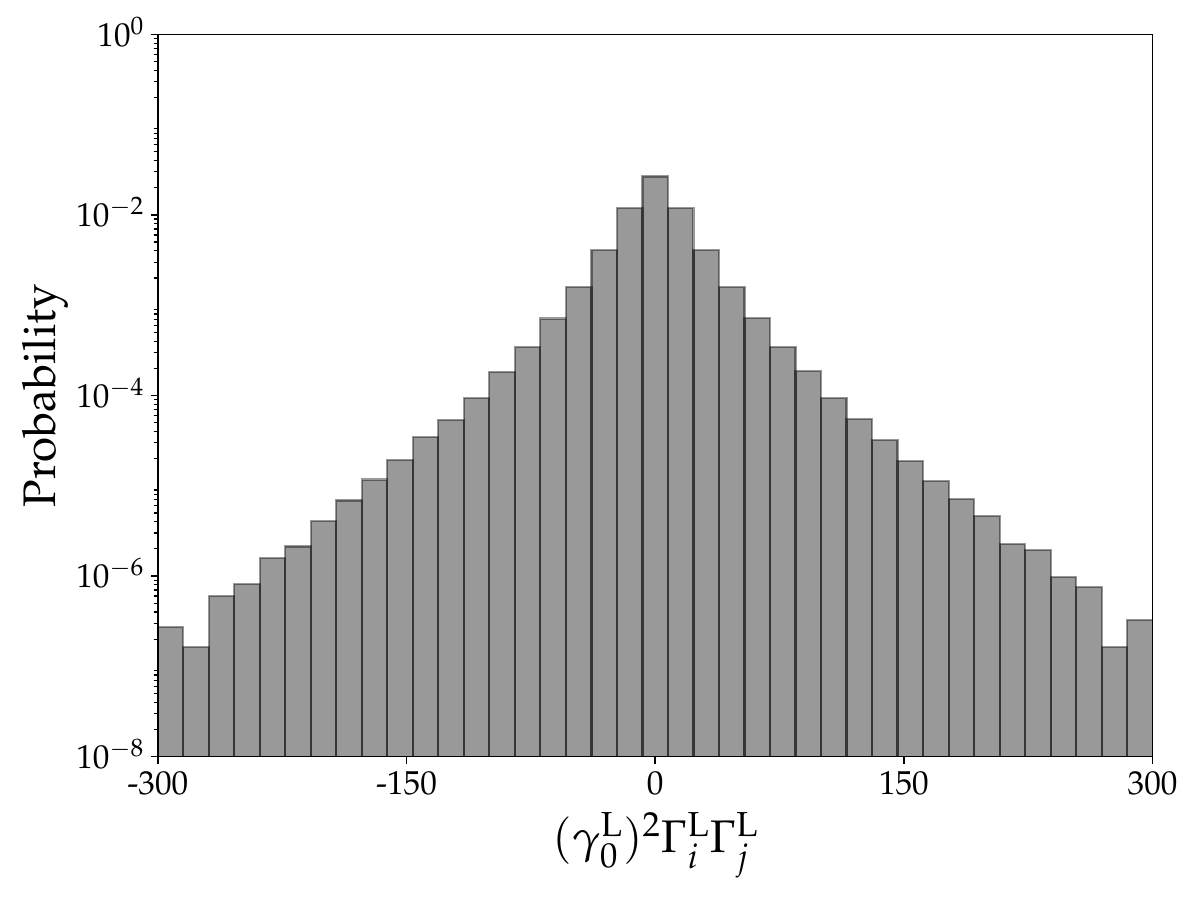}
	\caption{Histogram of the longitudinal squared deformation potential~\eqref{eq: Longitudinal deformation potential} in eV$^2$ for a single sample of a-TiO$_{2}$.}
	\label{fig: Gamma_square_distribution Ti}
\end{figure}

An example histogram of the product of longitudinal deformation potentials for one sample of a-Si is shown in Fig.~\ref{fig: Gamma_square_distribution} and one sample of TiO$_{2}$ in Fig.~\ref{fig: Gamma_square_distribution Ti}. States whose energy deforms in the same direction (increase or decrease energy) will have a positive product, and opposite directions a negative product. Similar to previous studies~\cite{levesque2022} we observe a fairly wide range of deformation potentials, with the product reaching up to $40 {\rm eV}^2$ in Fig.~~\ref{fig: Gamma_square_distribution}.

\section{Convergence tests}
\label{app: convtests}
\begin{figure}
	\includegraphics[width=\linewidth]{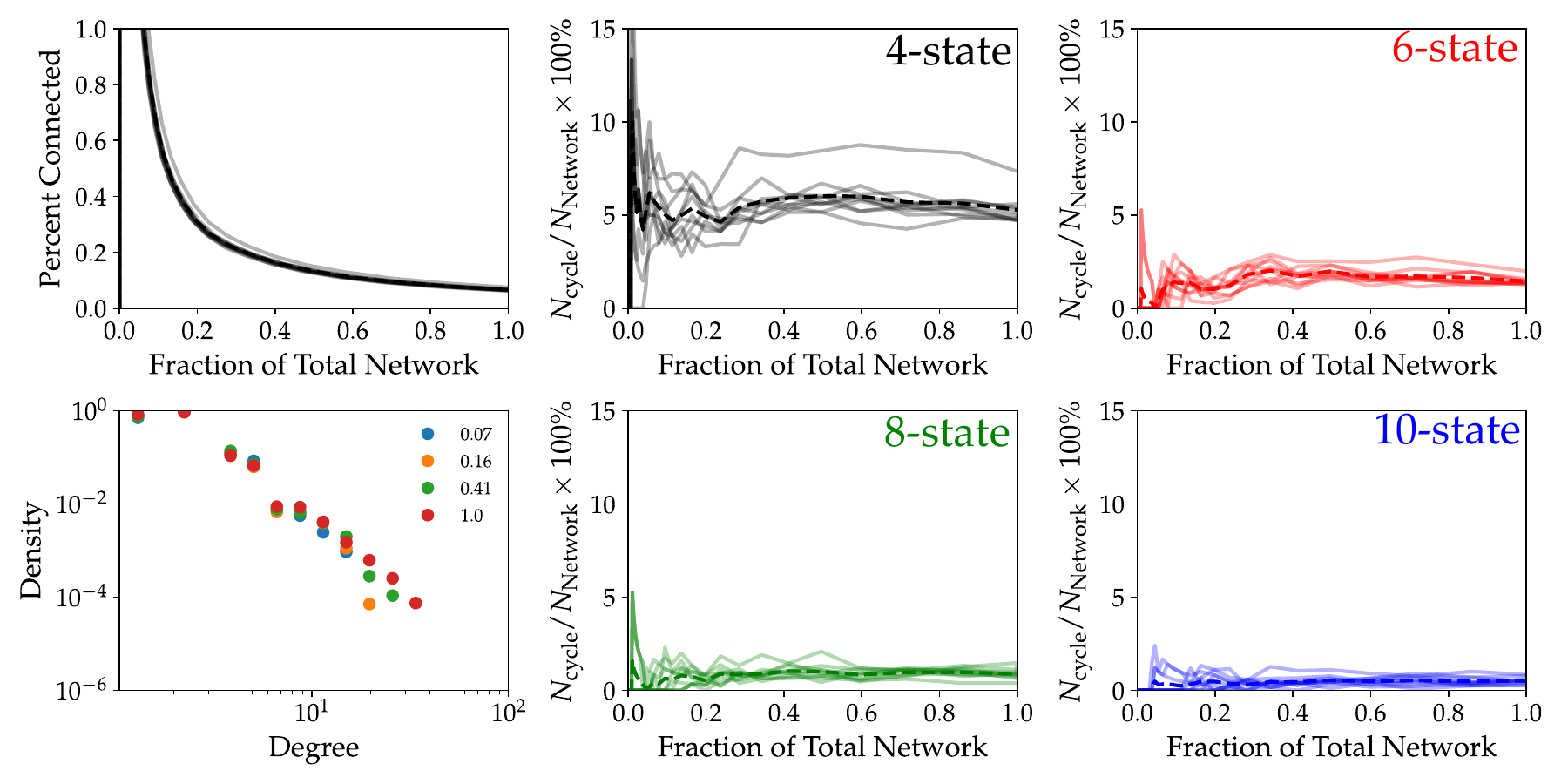}
	\caption{ Convergence tests for the statistical properties of our a-Si samples. The fraction of the total network is the number of subsampled nodes divided by the number of nodes in the full network. Each sample is shown as a faded solid curve with the average a completely opaque dashed curve. The percent of the network that is connected is defined as the number of connections divided by the maximum possible number of connections (all-all connected network). }
	\label{fig: convergence test Si}
\end{figure}

Finally, although our search for connected networks is by no means exhaustive, we have performed convergence tests for some key statistical properties and find them to be relatively robust. For a-Si (Fig.~\ref{fig: convergence test Si}) we find that the fraction of the network that is connected converges to $\sim 0.05 \%$ with relatively small spread between our 10 samples. The fraction of four-state cycles in the networks has a larger spread between samples, but remains relatively flat as the fraction of the total network included in the subsample approaches one. For higher order cycles, even faster convergence is observed. Comparing the degree distribution at logarithmically spaced fractions of the total network, we find that as the subsamples approach the full network, it always displays a scale-free structure (power law distribution).

\begin{figure}
	\includegraphics[width=\linewidth]{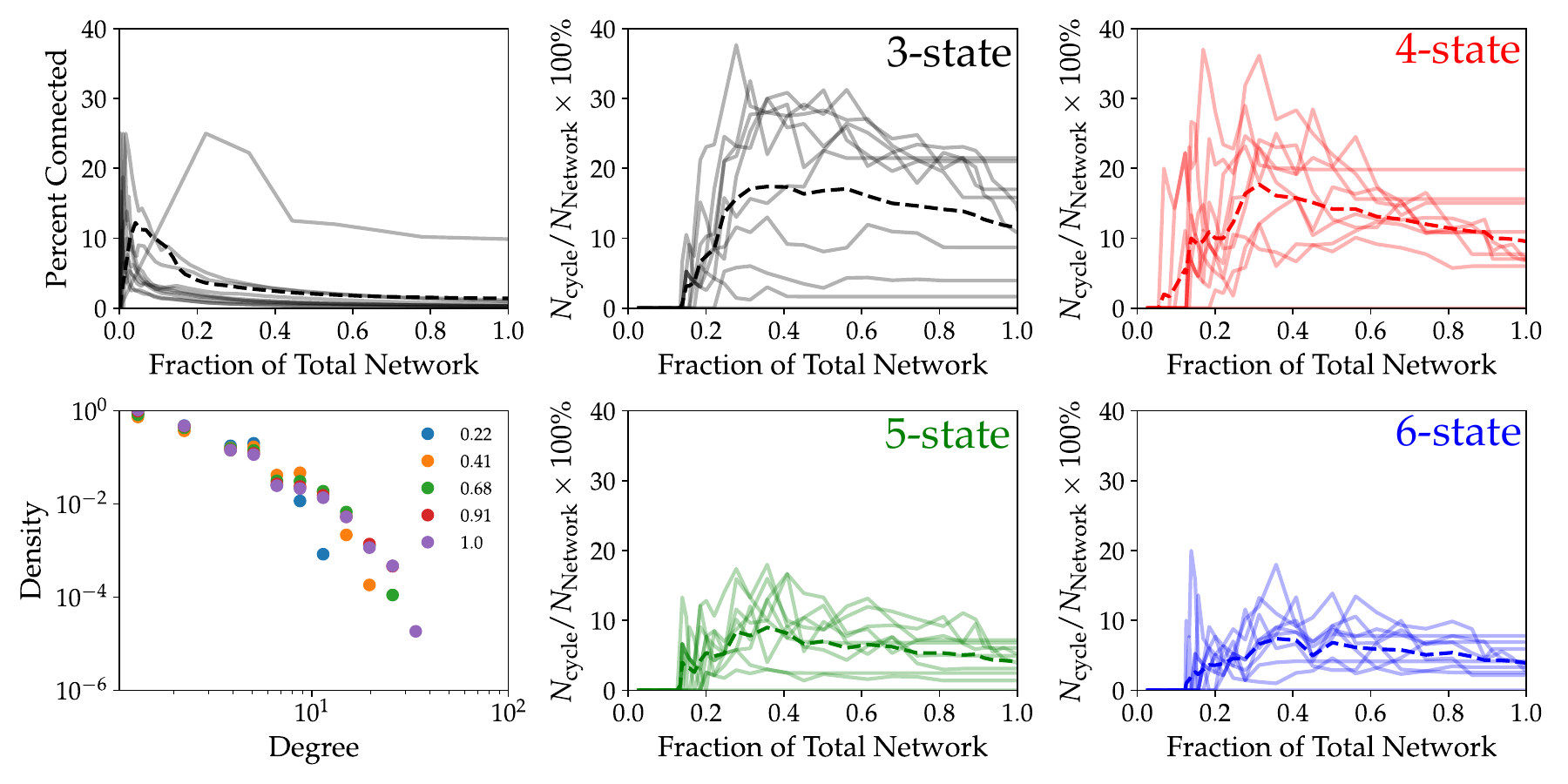}
	\caption{ Convergence tests for the statistical properties of a-TiO$_{2}$ samples, as in Fig.~\ref{fig: convergence test Si}.}
	\label{fig: convergence test Ti}
\end{figure}

Similar behavior is observed for the a-TiO$_{2}$ samples (Fig.~\ref{fig: convergence test Ti}); however, with significantly larger variation between samples and less stable convergence despite doubling the number of trajectories compared to a-Si. This is indicative of the sampling issues we faced. In general, we expect the qualitative trends to be robust to the incomplete sampling of the energy landscape and sampling biases, but quantitative predictions will likely change with improved sampling.

\end{document}